\documentclass[journal, final, onecolumn, 10pt]{IEEEtran}
\usepackage[nocompress]{cite}

\usepackage{subfigure}
\usepackage{float}
\usepackage{booktabs}

\usepackage[bookmarks=false, hidelinks]{hyperref}

\usepackage{amsmath,amssymb,amsfonts}
\allowdisplaybreaks
\usepackage{bm}                         % Bold calligraphic letters
\usepackage{multirow}
\usepackage{cases}                       % for numcases environment
\usepackage{tikz}
\usetikzlibrary{arrows.meta, positioning, automata, shapes, decorations.pathreplacing, calc, shadows.blur}

\usepackage[normalem]{ulem}             % strike text

\usepackage{amsthm}
\newtheorem{theorem}{Theorem}[section]
\newtheorem{lemma}[theorem]{Lemma}

\newtheorem{remark}{Remark}

\usepackage{graphicx}
\usepackage{textcomp}
\usepackage{xcolor}

\usepackage{balance}

\begin{document}

%[Stochastic Modeling and Resource Dimensioning of Multi-Cell Edge Intelligent Systems]
\title{Stochastic Modeling and Resource Dimensioning\\ of Multi-Cellular Edge Intelligent Systems
\thanks{}
}


\author{\IEEEauthorblockN{Jaume Anguera Peris, Joakim Jaldén}\\
\IEEEauthorblockA{School of Electrical Engineering and Computer Science\\
KTH Royal Institute of Technology, Stockholm, Sweden\\
Email: \{jaumeap, jalden\}@kth.se}
}

\maketitle

\begin{abstract}
Edge intelligence enables the execution of AI inference tasks on computing platforms at the network edge, typically co-located with or near the radio access network rather than in centralized clouds or on mobile devices. This approach is particularly well suited for data analytics of low-latency and resource-constrained applications, where large data volumes and stringent latency constraints require tight integration of wireless access and on-site computational resources. However, the performance and cost-efficiency of such systems fundamentally depend on the joint dimensioning of wireless and computational resources prior to deployment, specially amid spatial and temporal uncertainties. Prior works largely emphasizes run-time resource allocation or employs simplified network models that decouple radio access from computing infrastructure, overlooking end-to-end correlations in large-scale deployments. This paper introduces a unified stochastic framework for dimensioning multi-cellular edge-intelligent systems. We model network topology via a Poisson point process to capture randomness in user and base-station locations, incorporating inter-cell interference, distance-proportional fractional power control, and peak-power constraints. Integrating this with queueing theory and empirical profiling of AI inference workloads, we derive tractable expressions for the end-to-end offloading delay. These enable a non-convex joint optimization problem for minimizing deployment costs while enforcing statistical quality-of-service guarantees, defined not merely by averages, but by strict tail-latency and inference accuracy constraints. We prove decomposability into convex sub-problems, ensuring global optimality with zero gap. Through numerical evaluations in noise-limited and interference-limited regimes, we identify parameter regions that yield cost-efficient designs versus those that lead to severe under-utilization or unfairness across users. Key insights include the following: smaller cells reduce transmission delay but cause higher per-request computing cost due to reduced multiplexing at the servers, while larger cells exhibit the opposite trend. Moreover, network densification reduces computational costs only when frequency reuse scales with base-station density; otherwise, sparse deployments enhance fairness and efficiency in interference-limited scenarios. Overall, our analysis provides system designers with principled guidelines for scalable, QoS-aware provisioning of edge-intelligent video analytics in next-generation cellular networks.
\end{abstract}

\section{Introduction}
Edge intelligence has emerged as a paradigm shift in next-generation wireless networks, enabling the deployment of learning-based models on computing platforms at the network edge, typically co-located with or near the radio access network. This architecture is attractive for applications with latency stringent budgets and large data volumes, since it reduces reliance on centralized clouds or resource-constrained mobile devices, alleviates backhaul congestion, and enhances data privacy. For 5G and beyond-6G systems, these properties have made edge intelligence a key enabler for AI video analytics sector~\cite{li2023task}, with recent market projections indicating it is expected to grow from $15.15$ billion USD in $2025$ to $71.30$ billion USD by $2033$, at a compound annual growth rate of $21.4\%$, fueled by the expansion of IP-based surveillance in smart cities, industrial automation, and public safety infrastructure~\cite{grandview2025ai}. As such, realizing the full potential of edge-intelligent systems demands a precise dimensioning of the baseline resources (wireless spectrum and computational capacity) required to statistically guarantee performance before a network is even deployed.

Despite this critical need for network planning, the existing literature on edge video analytics has primarily concentrated on run-time optimizations. In this domain, where edge servers extract features to assist mobile users in real-time decision-making, mobile users benefit from edge intelligence if the gains of processing their images at the edge server are larger than the costs of transmitting the images through the wireless link. Consequently, significant progress has been made in operational strategies to maximize these gains, such as selecting adequate edge servers for fast response times~\cite{hamadi2022hybrid}, identifying and transferring only the most informative image regions~\cite{chen2022context}, adapting image formats to fluctuating network conditions~\cite{sun2022elasticedge}, or partitioning the deep leaning models between mobile users and edge servers to distribute image-processing workloads~\cite{khan2022distributed}. More recently, the work in~\cite{shokhanda2024safetail} has introduced a framework to manage computational redundancy and meet the strict tail-latency constraints required by safety-critical applications. However, while these works emphasize the increasing interest in edge video analytics, they only provide effective run-time strategies and do not address the dimensioning problem, which determines how many resources should be deployed a priori to ensure stability and cost-efficiency before these algorithms are even active.

Complementing these application-specific optimizations, a significant body of work addresses the operational phase more broadly, formulating the trade-off between network latency, computational latency, and accuracy. The work in~\cite{liu2018edge} proposed a multi-objective optimization problem to dynamically select the optimal edge server and video frame resolution. Extensions of this work account for the energy consumption of the entire offloading process, utilizing queuing theory to address the dynamic decisions in the system~\cite{wang2020joint}. Based on a similar idea, the authors in~\cite{wang2019effective} employed two-stage tandem queues to model transmission and computation, optimizing wireless and computing resources under statistical QoS guarantees. More recently, the work in~\cite{zhao2022edgeadaptor} further analyzed optimal resource provisioning for servers accommodating diverse deep learning models and configurations. These foundational works represented a critical advancement, moving from adapting parameters to understanding the cost-performance trade-offs of the physical infrastructure itself. However, they are still operational strategies that overlook the infrastructure dimensioning and planning stage, where the cost-efficiency of the system is determined by the initial deployment of spectral and hardware resources.

When shifting the focus to network planning and dimensioning, the challenge lies in modeling the inherent stochasticity of large-scale deployments. Network traffic is driven by spatially random user events, wireless channels are subject to path loss, fading, and interference, and edge servers face fluctuating inference workloads. Early works in this domain formulated multi-objective optimization problems to find optimal edge server locations to balance workloads and minimize latency~\cite{wang2019edge}. Besides, the work in~\cite{ko2018wireless} analyzed the trade-off between the average computation latency and the network connectivity in a more realistic scenario by considering the inherent aspects of noise, fading, and interference in wireless communications. This was later extended in~\cite{mukherjee2020edge} to minimize energy consumption for statistical guarantees of the joint success of wireless transmission and task computation. With this latter work, the integration of stochastic geometry with edge intelligence has gained traction. The work in~\cite{li2023task} derived performance metrics for task-oriented heterogeneous networks, emphasizing the need for semantic-aware resource planning. In the context of vehicular networks, the work in~\cite{yang2024game} applied stochastic geometry to model computation offloading under high mobility and interference. Furthermore, the authors in~\cite{saeedi2025stochastic} utilized stochastic geometry to model partial offloading reliability in distributed edge-AI systems.

Despite these significant advances in joint communication and computation schemes, a critical gap remains in the literature regarding the dimensioning of large-scale infrastructure. Existing single-cell frameworks predominantly optimize resources under the assumption of homogeneous network conditions, failing to capture the spatial randomness inherent in realistic deployments. On the other hand, while advanced multi-cellular systems incorporate the stochastic nature of wireless communication, they face two distinct limitations regarding infrastructure dimensioning. First, many frameworks decouple resource planning: radio access networks are dimensioned based on coverage capabilities or spectral efficiency metrics, while computing infrastructure are sized using independent queueing approximations. This isolation fails to capture the end-to-end joint trade-off, where a spectral bottleneck can induce upstream starvation at compute nodes, while under-provisioned servers with excessive queuing or inference delays can render high-capacity wireless links ineffective. Second, even when resources are dimensioned jointly, existing approaches predominantly rely on average performance metrics to ensure tractable convex optimization problems, failing to address the strict tail-latency constraints required by next-generation applications.

To bridge this gap, we introduce a unified stochastic modeling framework for resource dimensioning in multi-cellular edge-intelligent systems. By integrating stochastic geometry, queueing theory, and statistical modeling of AI inference workloads, we model the end-to-end performance of edge-intelligent systems within a spatially random network topology. While the framework is designed to be versatile and generalizable to various edge-intelligent workloads, we illustrate the key concepts through the lens of edge video analytics as a representative high-bandwidth and latency-critical use case. This allows us to formulate a joint optimization problem that determines the minimum cost-efficient bandwidth and computing capacity required to satisfy statistical quality-of-service guarantees, defined not merely by averages, but by strict tail-latency and inference accuracy constraints.

\subsection{Our contribution}
This paper addresses the resource-dimensioning problem for a multi-cell system supporting edge video analytics through rigorous performance modeling and analytical evaluation. In particular, the novelty of this work lies in the development an optimization framework that enable precise performance analysis and evaluation of computing systems under uncertainty, as follows:

\begin{itemize}
    \item We present a stochastic geometry framework for the network topology of a single-input multiple-output (SIMO) communication system, derive closed-form expressions for the ergodic capacity of the uplink transmission in both noise-limited and interference-limited regimes, and provide design guidelines regarding key network parameters, such as base station density, frequency reuse factor, and power control coefficient.
    
    \item We characterize the dynamics of the task offloading process for both the noise-limited and interference-limited systems using queueing theory, and we model the inherent relationship between the bandwidth and the computational resources as a function of the supported arrival rate of tasks at the edge server, the maximum permitted end-to-end delay in the system, and the density of base stations in the network.

    \item We formulate a non-convex optimization problem to jointly optimize the communication and computing resources such that a given set of users and a given set of edge servers meet the accuracy constraints of the video analytics and satisfy a minimum probability of successfully completing the end-to-end process within a delay requirement.

    \item We break down the non-convex optimization problem into a series of convex sub-problems and provide a theoretical analysis of the guarantees under which the sequential evaluation of these sub-problems leads to a globally optimal solution of the large-scale resource-dimensioning problem.
    
    \item We evaluate the optimal solution of the non-convex optimization problem through simulation-based analysis and discuss the trade-off between the different parameters of the system, with special focus on how the system scales for varying density of base stations in the network, traffic intensity, per-user power control settings, and cellular reuse factors.
\end{itemize}

By the end of this paper, readers will have a comprehensive understanding of the importance of resource dimensioning for edge-intelligent, multi-cellular systems supporting video analytics, and the key factors that must be considered for the joint optimization of wireless and computing resources under statistical QoS constraints.

\subsection{Document organization}
\label{subsec:document_organization}
The remainder of this paper is organized as follows. Section~\ref{sec:network_model} presents the stochastic geometry framework for the network topology, derives the closed-form ergodic capacity for the uplink transmission, and provides design guidelines regarding key network parameters. Section~\ref{sec:offloading_model} characterizes the end-to-end latency, modeling the frame arrival process, edge-server queueing dynamics, and the AI inference workload. In Section~\ref{sec:resource_dimensioning}, we formulate the joint resource-dimensioning optimization problem, define the statistical quality-of-service requirements, and provide a theoretical analysis of the global optimality of the solution. Section~\ref{sec:numerical_results} presents the numerical evaluation of the proposed strategies in both noise-limited and interference-limited regimes. Finally, we summarize the main findings and the concluding remarks in Section~\ref{sec:conclusions}.

\section{Network model}
\label{sec:network_model}
\subsection{Base station deployment and user association}
Consider a large-scale cellular network where the location of the base stations are modelled as a two-dimensional homogeneous Poisson point process (PPP) with intensity $\lambda_{\mathrm{b}}$. This stochastic geometry approach captures the spatial randomness inherent in real-world cellular deployments, enabling tractable analysis of key network parameters while aligning with established empirical observations~\cite{chiu2013stochastic}. As a direct consequence of this PPP model, we derive three fundamental parameters of interest, illustrated in Figure 2(a), which characterize the geometric relationships between users and base stations.

The first parameter of interest is the distance $r$ between a typical user and its nearest serving base station, a critical factor in determining coverage and signal quality. The cumulative distribution function (CDF) of the normalized distance, defined as $\bar{r} = \sqrt{\lambda_{\mathrm{b}}} r$, can be derived from the null probability of a two-dimensional PPP and follows a Rayleigh distribution:
\begin{equation}
    \mathcal{P}(\bar{r}\leq x) = 1 - e^{-\pi x^2}, \quad x\geq 0.
    \label{eq:cdf_normalized_distance}
\end{equation}
This distribution captures the probabilistic nature of user-base station associations, providing a foundation for analyzing the reliability of uplink transmissions.

The second parameter on interest is the maximum distance $r_{\max}$ from any user in a cell to its serving base station, which defines the extent of Voronoi cells and informs of coverage limits for cell-edge users. For the normalized maximum distance $\bar{r}_{\max} = \sqrt{\lambda_{\mathrm{b}}} r_{\max}$, extensive numerical studies in the literature have established that its CDF is well-represented by a Gamma distribution~\cite{peris2024extreme}:
\begin{equation}
    \mathcal{P}(\bar{r}_{\max}\leq x\mid \alpha, \beta, \gamma) = \frac{\alpha \beta^{\gamma/\alpha}}{\Gamma(\gamma/\alpha)} \int_0^x t^{\gamma-1} e^{-\beta t^\alpha} \, dt, \quad x \geq 0,
    \label{eq:cdf_normalized_maxDistance}
\end{equation}
with parameters $(\alpha, \beta, \gamma) = (1.719, 5.528, 9.482)$, where $\Gamma(z) = \int_0^{\infty} t^{z-1}e^{-t}dt$ is the Gamma function. This form captures the tail behavior of cell radii in random deployments, ensuring accurate modeling of the worst-case user experience.

Lastly, the third parameter of interest is the area $A$ of the Voronoi cells formed by the base stations, which governs the spatial distribution of traffic and computational load across the network. For the normalized cell area $\bar{A} = \lambda_{\mathrm{b}} A$, prior analyses have established that its CDF is effectively approximated by a Gamma distribution~\cite{ferenc2007size}:
\begin{equation}
    \mathcal{P}(\bar{A} \leq x\mid \alpha, \beta, \gamma) = \frac{\alpha \beta^{\gamma/\alpha}}{\Gamma(\gamma/\alpha)} \int_0^x t^{\gamma-1} e^{-\beta t^\alpha} \, dt, \quad x \geq 0,
    \label{eq:cdf_normalized_area}
\end{equation}
with parameters $(\alpha, \beta, \gamma) = (1, 3.5, 3.5)$. This distribution accounts for the variability in cell sizes, enabling realistic modeling of traffic distribution for different network densities.

Note that these three measures ($r$, $r_{\max}$, and $A$) all scale inversely to the density of base station per unit area for any arbitrary $\lambda_{\mathrm{b}}$ > 0, yet remain independent of other system variable. Besides, these geometric characterizations form the basis for modeling uplink transmissions in the multi-cell environment, where user distances and cell areas directly influence signal quality and interference, as explored next.

\subsection{Uplink transmission model}
\label{sec:system-model_ULtransmission}
Consider a large-scale multi-cell SIMO network where each base station is equipped with $M \geq 1$ antennas and is co-located with an edge server supporting AI-based applications for video analytics. Mobile users are uniformly distributed over the network area and are associated with the base station that provides the maximum received power averaged over fading. Each video-analytics user is allocated a dedicated bandwidth $B$ for uplink frame transmission through a high-priority dynamic network slice that guarantees immediate wireless access~\cite{rost2017network}. Within each slice, base stations employ frequency-division multiple access with frequency reuse factor $\delta$, resulting in inter-cell interference but eliminating intra-cell interference. Furthermore, each base station schedules exactly one active user per frequency sub-band, selected uniformly at random from the users within its Voronoi cell.
\begin{figure*}
    \centering
    \subfigure[]{\includegraphics[width = 0.45\textwidth]{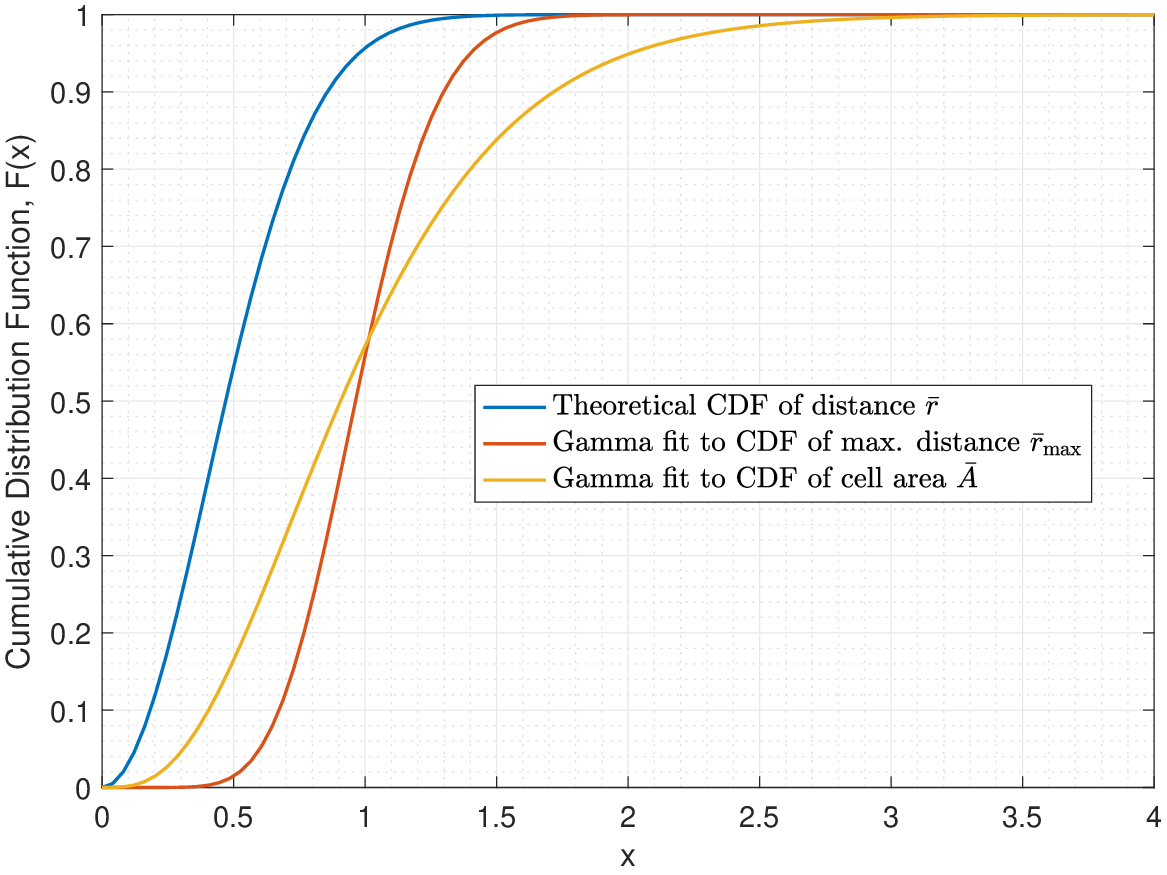} \label{fig:CDF_distributions}}\hspace{.4cm}
    \subfigure[]{\includegraphics[width = 0.45\textwidth]{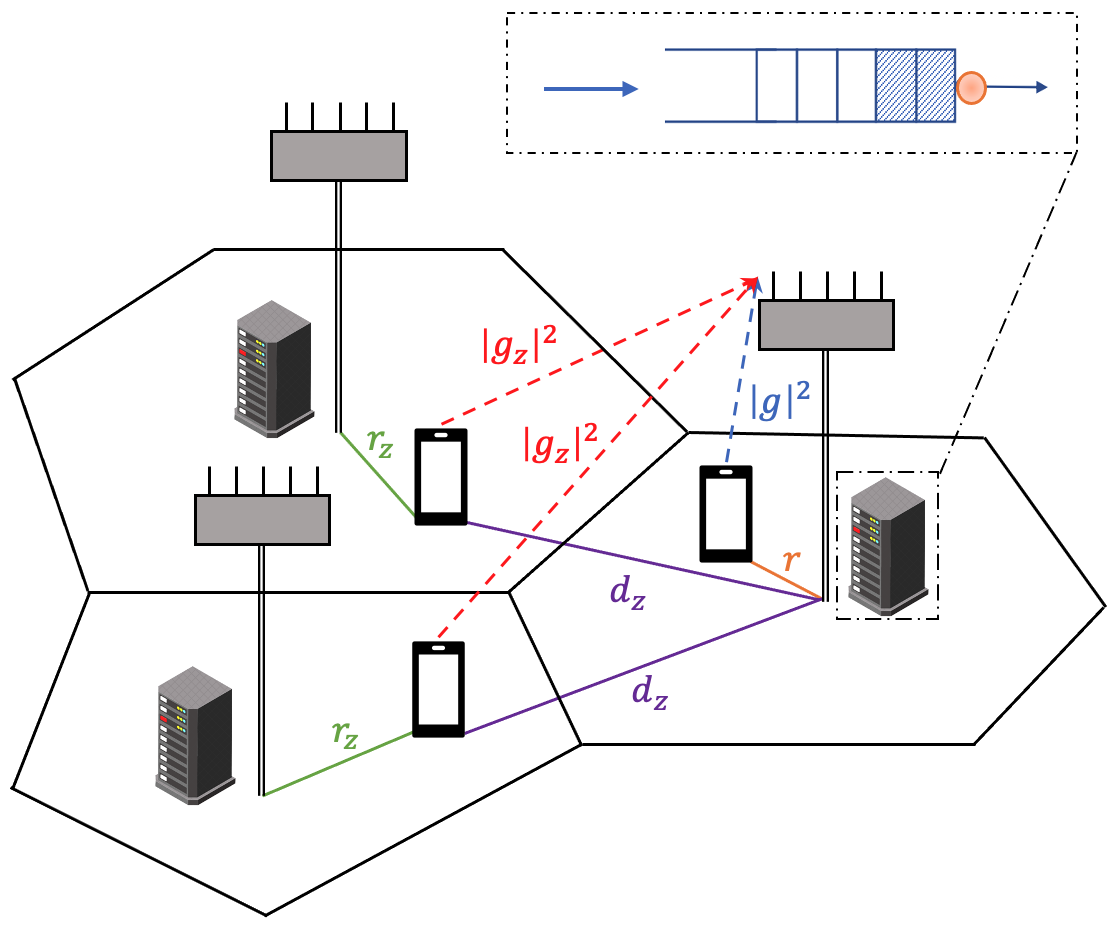} \label{fig:interferenceModel}}
    \caption{(a) Cumulative distribution function of the normalized distance between any user and its associated base station, and best-fit approximations to the cumulative distribution functions of the normalized maximum distance squared \eqref{eq:cdf_normalized_maxDistance} and the normalized area of the Voronoi cells in two dimensions \eqref{eq:cdf_normalized_area}, taken from~\cite{peris2024extreme} and~\cite{ferenc2007size}, respectively. (b) Multi-cell system with reuse factor $\delta=1$, where the user of interest is located at the Voronoi cell of the base station on the right, and the interfering users are located elsewhere. The straight lines show the different distances $r$, $r_z$, and $d_z$ defined in Section~\ref{sec:system-model_ULtransmission}. The dashed lines show the signal of interest (blue) and the interfering signals (red) at one of the $M$ antennas at the receiver, characterized by the different Rayleigh fading coefficients of the channel $|g|^2$, and $|g_z|^2$, respectively.}
    \label{fig:networkModel}
\end{figure*}

The signals between any pair of transmitter and receiver antennas experience independent and identically distributed (i.i.d.) Rayleigh fading with path-loss exponent $\alpha > 2$. The effective scalar channel gain after maximum-ratio combining at the base station is modeled as $\|\bm{g}\|^2 = \sum_{m=1}^M |g_m|^2$, where $|g_m|^2 \sim \exp(\gamma)$ is an Exponential random variable representing the small-scale fading, $\gamma = (\lambda_c / (4\pi))^2$ denotes the distance-dependent path loss, and $\lambda_c$ is the wavelength of the carrier frequency. Consequently, the channel gain $\|\bm{g}\|^2$ for $M$ co-located antennas follows a Gamma distributed with shape parameter $M$ and scale parameter $\gamma$.

Users implement distance-proportional fractional power control of the form $P r^{\alpha \epsilon}$, where $P$ is the reference power at 1 kilometer, $r$ is the distance to its serving base station, and $\epsilon \in [0, 1]$ is the power control coefficient. To account for the finite peak power of mobile devices, the transmit power is capped by the maximum transmit power $\bar{P}$. The resulting distance-dependent transmit power function, incorporating both fractional power control and the peak-power constraint, is therefore defined as
\begin{equation}
    \ell(r, \alpha, \epsilon) = \min(P r^{\alpha \epsilon}, \bar{P}).
    \label{eq:peakPowerConstrain}
\end{equation}
The additive receiver noise is white Gaussian with power $\sigma^2$.

Inter-cell interference arises from the set $\mathcal{Z}$ of co-channel users active in other cells. Let $r_z$ denote the distance from interferer $z \in \mathcal{Z}$ to its serving base station, and $d_z$ the distance from interferer $z$ to the base station of interest (see Figure~\ref{fig:interferenceModel}). Note that the distances $\{r_z\}_{z \in \mathcal{Z}}$ are identically distributed but not necessarily independent. The dependence arises from the structure of the Poisson-Voronoi tessellation and the single-user-per-band scheduling.  However,~\cite{novlan2013analytical} demonstrates that this dependence is weak, allowing the distances $r_z$ to be approximated as i.i.d. As such, and following the same null-probability argument as for the desired link distance $r$, we can model the marginal distribution of $r_z$ as Rayleigh with cumulative distribution function given in~\eqref{eq:cdf_normalized_distance}. Lastly, we consider the interfering signals to be projected onto the beamforming space in the direction of the signal of interest via $\Pi_{\bm{g}} = \bm{g} \bm{g}^H / \|\bm{g}\|^2$. The resulting effective interference channel gain is $\|\Pi_{\bm{g}} \bm{g}_z\|^2  \sim \exp(\gamma)$ due to the isotropic nature of Rayleigh fading in the projected subspace.

Under this transmission model, the uplink signal-to-interference-plus-noise ratio (SINR) for a typical user at distance $r$ from its serving base station is:
\begin{equation}
    \text{SINR} = \frac{\|\bm{g}\|^2_2 \,\ell(r,\alpha,\epsilon)\,r^{-\alpha}}{\sigma^2 +\sum_{z\in\mathcal{Z}} \|\Pi_{\bm{g}}\,\bm{g}_z\|^2_2\, \ell(r_z,\alpha,\epsilon)\, d_z^{-\alpha}}.
    \label{eq:SINR}
\end{equation}

This SINR expression lays the foundation for analyzing the trade-offs between spectral resources, base-station density $\lambda_{\mathrm{b}}$, and reuse factor $\delta$. To evaluate the performance of the uplink transmission, we adopt the ergodic capacity of the channel as the achievable uplink transmission rate, reflecting the system's ability to enable fast or low-power transmission for video-analytics users. In particular, depending on the different network parameters, the system may operate in a noise-limited or interference-limited regime. We derive next the ergodic capacity for these cases separately, for a user with allocated bandwidth $B$ transmitting at a distance $r$ from the base station.

\begin{lemma}[Noise-limited system]
\label{lemma:noise_limited}
    Consider a noise-limited uplink transmission in a single-input multiple-output system with $M \geq 1$ receiver antennas at the base station. Let a user at distance $r$ (in kilometers) from its serving base station transmit with power $\ell(r, \alpha, \epsilon) = \min(P r^{\alpha \epsilon}, \bar{P})$ over a bandwidth $B$ (in Hz). Moreover, let the small-scale fading channel gain $\|\bm{g}\|^2$ follow a Gamma distribution with shape parameter $M$ and scale parameter $\gamma$, as established in the channel model, and let the noise power spectral density be $N_0 = \sigma^2 / B$, where $\sigma^2$ is the receiver noise power. Then, the ergodic capacity (in bits per second) is given by
    \begin{equation}
        C_{\mathrm{NL}}(B, r) = \frac{B}{\log(2)} \, \exp\left(\frac{BN_0 \,r^\alpha}{\gamma \ell(r,\alpha,\epsilon)} \right) \sum_{i=0}^{M-1} E_{i+1}\left(\frac{BN_0\, r^\alpha}{\gamma \ell(r,\alpha,\epsilon)} \right)
        \label{eq:ergodicCapacity_noiseLimited}
    \end{equation}
    where $E_i(x) = \int_1^\infty \exp(-xt)/t^i\, dt$ is the generalized exponential integral function for $i \geq 1$.
    \begin{proof}
        Appendix~\ref{app:erg_NL}.
    \end{proof}
\end{lemma}

\begin{lemma}[Interference-limited system]
\label{lemma:interference_limited}
    Consider an interference-limited uplink transmission in a multi-cell single-input multiple-output system with $M \geq 1$ receiver antennas at the base station, operating under frequency-division multiple access with reuse factor $\delta$. Let a user at distance $r$ (in kilometers) from its serving base station transmit with power $\ell(r, \alpha, \epsilon) = \min(P r^{\alpha \epsilon}, \bar{P})$ over a bandwidth $B$ (in Hz). Let the small-scale fading channel gain $\|\bm{g}\|^2$ follow a Gamma distribution with shape parameter $M$ and scale parameter $\gamma$, and let the maximum-ratio combining vector for beamforming projection onto the channel of interest be $\Pi_{\bm{g}} = \bm{g} \bm{g}^H / \|\bm{g}\|^2$. Let base stations be distributed according to a homogeneous Poisson point process with density $\lambda_{\mathrm{b}}$ (base stations per unit area), and let inter-cell interference arise from a set $\mathcal{Z}$ of active co-channel users, each with distance $d_z$ to the base station of interest and distance $r_z$ to their serving base station, where $\{r_z\}_{z \in \mathcal{Z}}$ are i.i.d. Rayleigh distributed. Then, the ergodic capacity (in bits per second) is given by
    \begin{equation}
        C_{\mathrm{IL}}(B, r) = \frac{B}{\log(2)} \int_{0}^{\infty} \Big(1-\frac{1}{(1+s\gamma \ell(r,\alpha,\epsilon)\,r^{-\alpha})^{M}}\Big) \frac{\mathcal{L}(s)}{s}\, ds,
        \label{eq:ergodicCapacity_interferenceLimited}
    \end{equation}
    where the Laplace transform of the interference is
    \begin{equation}
        \mathcal{L}(s) = \exp\left(-2\pi\frac{\lambda_{\mathrm{b}}}{\delta}\int_{r}^{\infty} \beta(x, s)\, x\, dx  \right), 
        \label{eq:laplaceTransform_interference}
    \end{equation}
    with
    \begin{equation}
        \beta(x, s) = 1-\int_{0}^{\infty}\frac{2\pi\lambda_{\mathrm{b}}u\, e^{-\pi\lambda_{\mathrm{b}} u^2}}{1+s\gamma \ell(u,\alpha,\epsilon) x^{-\alpha} } \,du.
    \end{equation}
    \begin{proof}
        Appendix~\ref{app:erg_IL}.
    \end{proof}
\end{lemma}

For convenience, Table~\ref{tab:sysparams} summarizes the system parameters and notation used throughout the network, offloading, and inference models. Scenario-dependent parameters and baseline parameter values are specified in Section~\ref{sec:numerical_results}.

\begin{table*}
\caption{System parameters and notation for the multi-cell edge video analytics model}
\label{tab:sysparams}
\centering
\begin{tabular}{@{}lp{0.56\linewidth}@{}}
\toprule
\textbf{Symbol} & \textbf{Description} \\
\midrule
\multicolumn{2}{@{}l@{}}{\emph{Network geometry and spectrum}}\\
$\lambda_b$ & Spatial density of base stations in the network \\
$\delta$ & Number of orthogonal sub-bands for inter-cell interference mitigation \\
$\alpha$ & Power-law exponent of the large-scale path-loss model \\
$f_c$ & Center frequency of the uplink transmission band \\
\midrule
\multicolumn{2}{@{}l@{}}{\emph{Uplink transmission and power control}}\\
$\epsilon$ & Fractional compensation of path loss in uplink power control \\
$P$ & Nominal transmit power normalized to unit distance \\
$\bar{P}$ & Maximum allowable uplink transmit power per user \\
$N_0$ & Noise power spectral density at the receiver \\
$M$ & Number of receive antennas at each base station \\
\midrule
\multicolumn{2}{@{}l@{}}{\emph{Video frame and traffic model}}\\
$s^2$ & Frame resolution of a video frame (pixels) \\
$\theta$ & Number of bits per pixel after encoding \\
$\xi$ & Video compression factor applied before transmission \\
$\lambda$ & Frame arrival intensity of video frame generation requests \\
\midrule
\multicolumn{2}{@{}l@{}}{\emph{Edge inference and computing model}}\\
$H$ & Total available computing resources at an edge server \\
$H_f$ & Computing resources allocated per processed frame \\
$T_s$ & Inference processing time for a video frame \\
$a(s)$ & Accuracy of the AI model as a function of frame size \\
$c_1,c_2$ & Latency model constants of the inference service-time model \\
$c_3,c_4,c_5$ & Accuracy model constants of the inference accuracy model \\
\midrule
\multicolumn{2}{@{}l@{}}{\emph{QoS constraints and optimization}}\\
$D$ & End-to-end latency constraint for offloading and inference \\
$\rho$ & Utilization factor of the edge server queue \\
$\omega_{\min}$ & Minimum success probability of meeting the delay constraint \\
$\eta_r$ & Cell-edge coverage (reliability constraint for worst-case user) \\
$\eta_A$ & Cell-area coverage (reliability constraint over large Voronoi cells) \\
$a_{\min}$ & Minimum inference accuracy \\
$\beta_1,\beta_2$ & Cost weights of wireless and computing costs \\
$\vartheta$ & Regularization parameter in the optimization problem \\
\bottomrule
\end{tabular}
\end{table*}

\subsection{Performance insights and design guidelines}
The closed-form ergodic capacities derived in Lemmas~\ref{lemma:noise_limited} and~\ref{lemma:interference_limited} enable a systematic exploration of uplink performance in edge-enabled multi-cell networks. While some trends align with intuition, several new insights that are not readily available in prior stochastic-geometry literature emerge due to the combined consideration of fractional power control with peak-power constraint, dynamic frequency reuse, and realistic distance distributions in Poisson--Voronoi cells. With that in mind, we now evaluate their implications through numerical studies, uncovering trade-offs in the different network design parameters.

From the analysis in Section~\ref{sec:system-model_ULtransmission}, notice that the power control coefficient $\epsilon$ plays and important role on the transmitted power. When $\epsilon = 0$, the fractional power control provides an energy-preserving solution in which all users transmit at the same power, and when $\epsilon = 1$, the fractional power control aims to compensate for the propagation losses, but, because of that, it is easier that users further from their serving base station reach their peak power constraint. At the same time, since we are considering that mobile users operate with isotropic antennas, higher transmitted powers also lead to potentially higher interference. The two most important parameters to control the interference in our model are the density of base stations $\lambda_{\mathrm{b}}$ and the reuse factor $\delta$.
%, which come into play into the resource-dimensioning problem via \eqref{eq:laplaceTransform_interference}.
For our specific case scenario, we consider $\lambda_{\mathrm{b}}$ to determine the density of base stations per unit area, and consequently, the density of edge servers in the network, and consider $\delta$ to determine the number of different frequency bands used in the network, with just one band assigned per base station. Hence, to aim for a fair comparison, the ratio $\lambda_{\mathrm{b}}/\delta$ must remain constant, as this leads to a system in which the total number of frequency bands, and hence interferers, remain the same, and the only thing that changes is the way in which these bands are distributed over the network. Considering this, we analyze the noise-limited and interference-limited scenarios separately.

\begin{figure}
    \centering
    \subfigure[]{\includegraphics[width=0.44\textwidth]{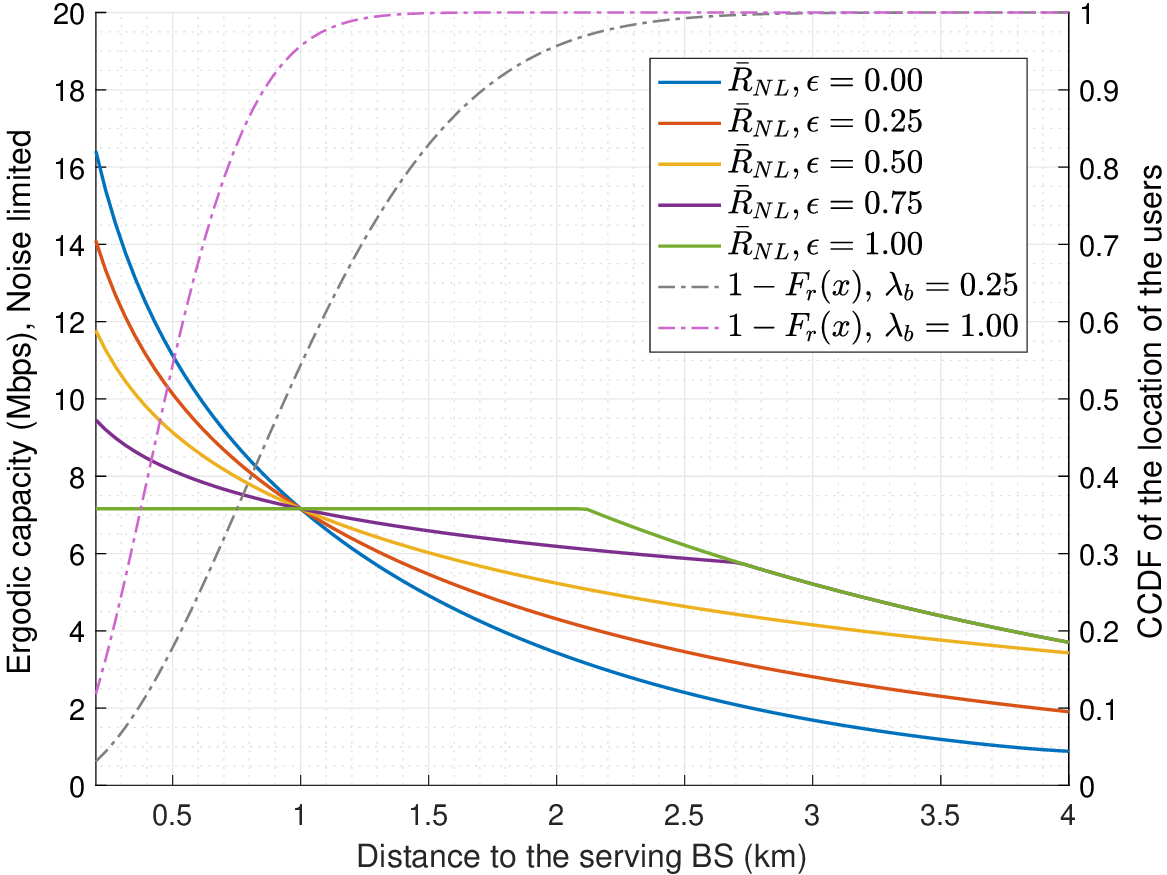} \label{fig:ergCapNL_sameM}}$\quad$
    \subfigure[]{\includegraphics[width=0.44\textwidth]{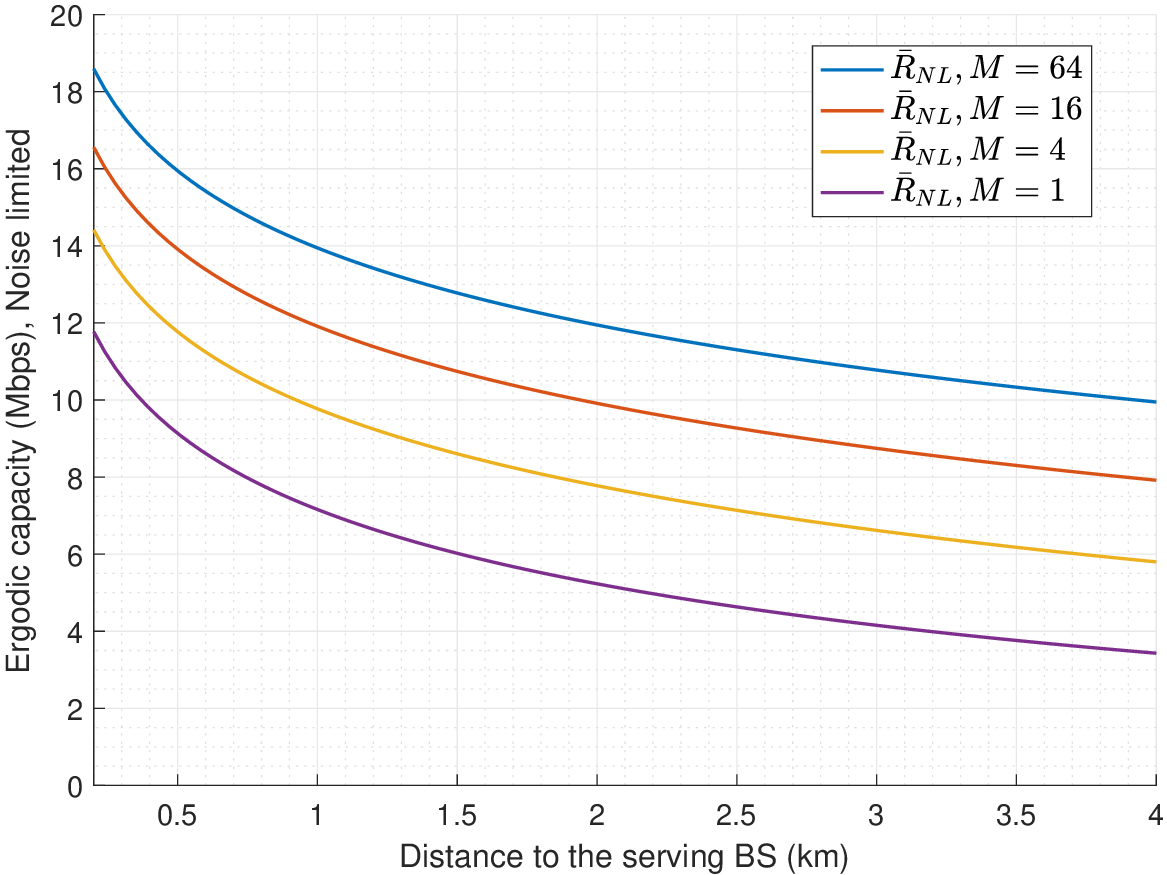} \label{fig:ergCapNL_sameEpsilon}}
    \subfigure[]{\includegraphics[width=0.44\textwidth]{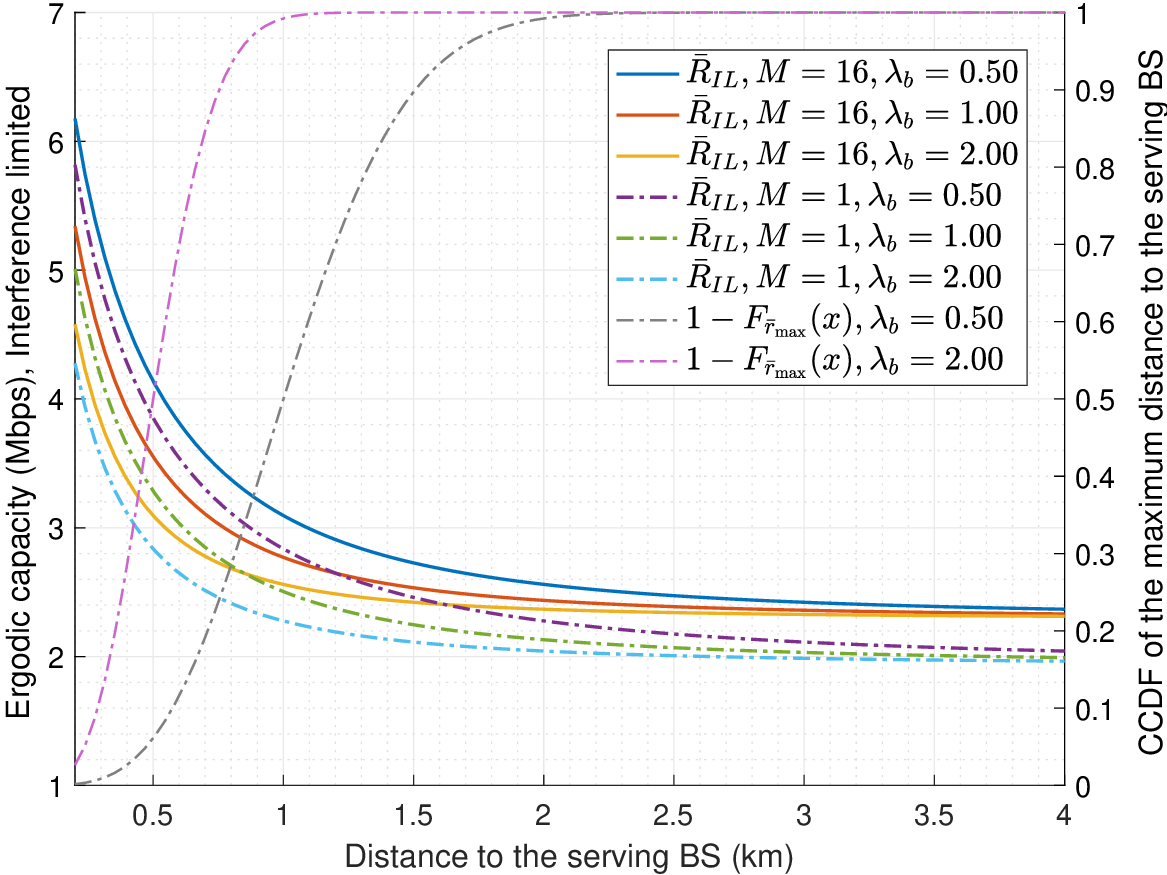} \label{fig:ergCapIL_diffRatio_delta4}}$\quad$
    \subfigure[]{\includegraphics[width=0.44\textwidth]{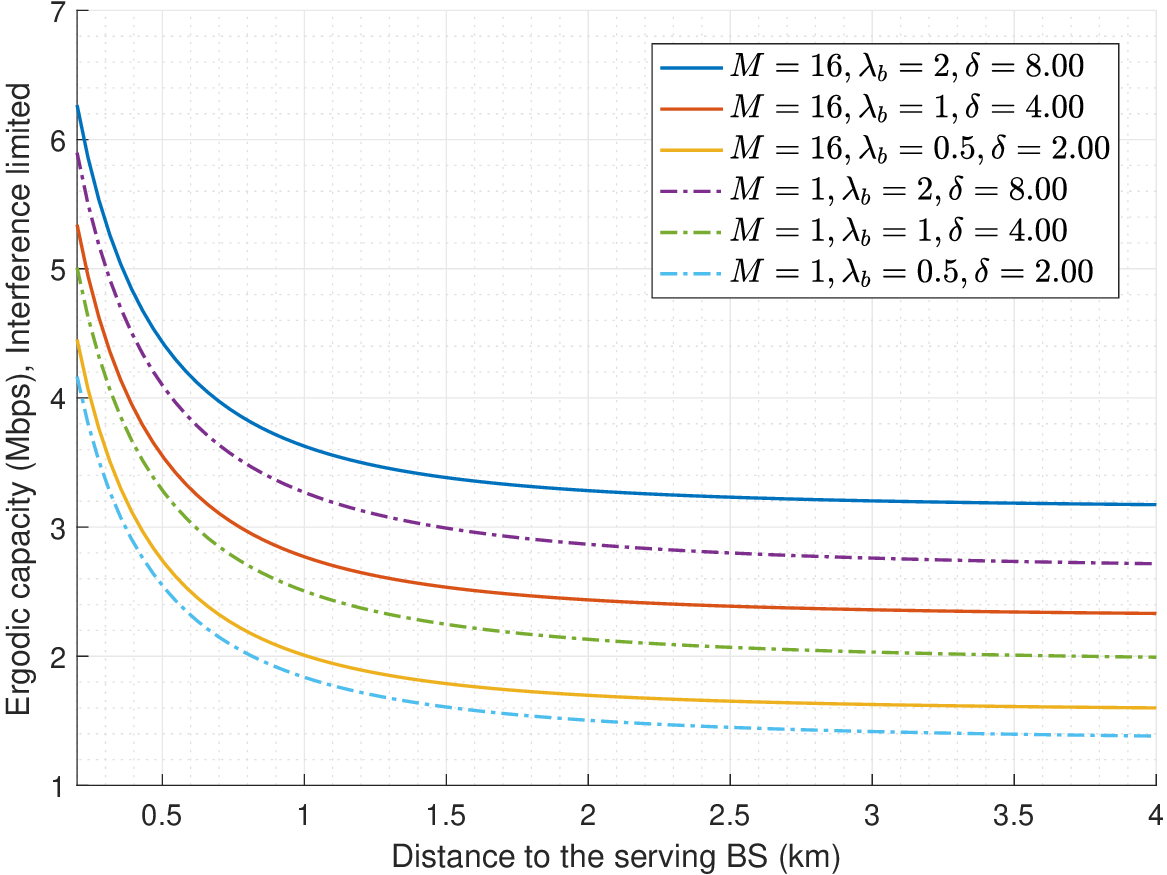} \label{fig:ergCapIL_sameRatio_ratio025}}
    \subfigure[]{\includegraphics[width=0.44\textwidth]{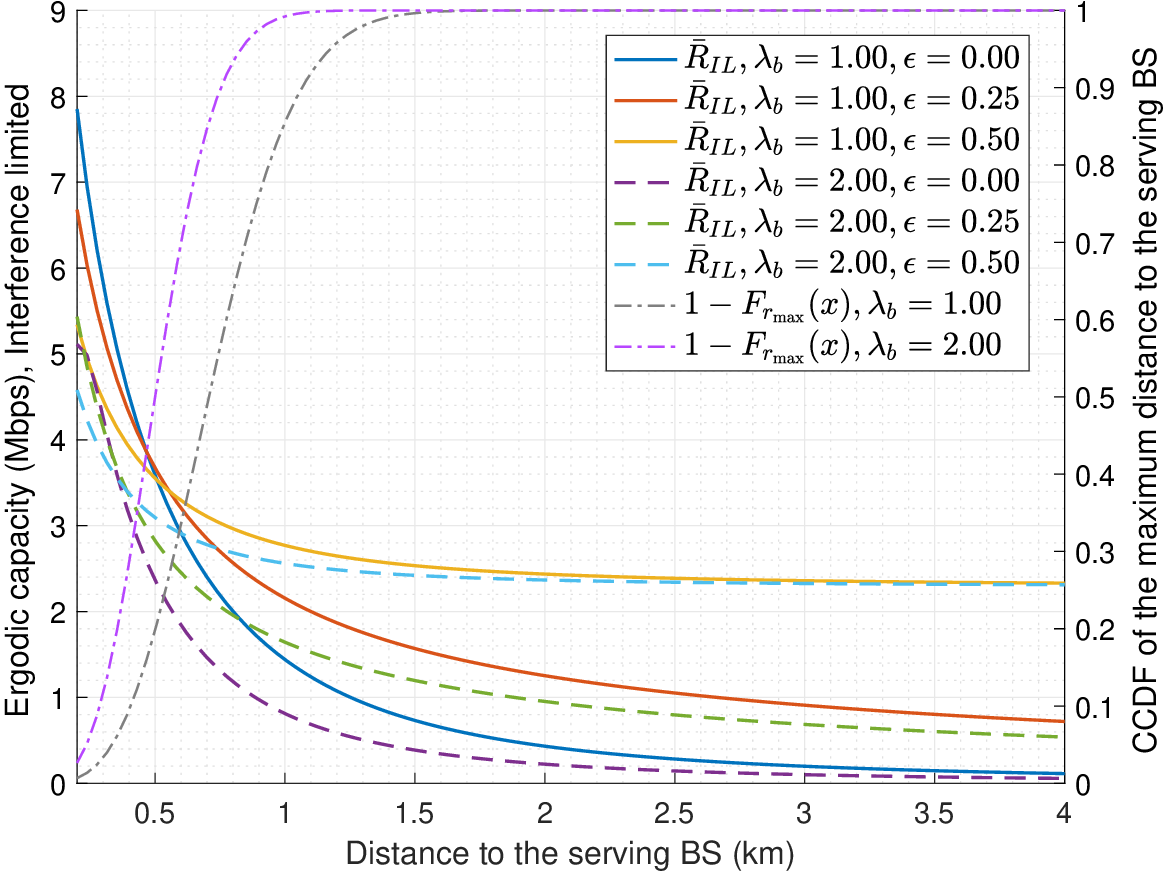} \label{fig:ergCapIL_diffRatio_delta4_epsilon}}$\quad$
    \subfigure[]{\includegraphics[width=0.44\textwidth]{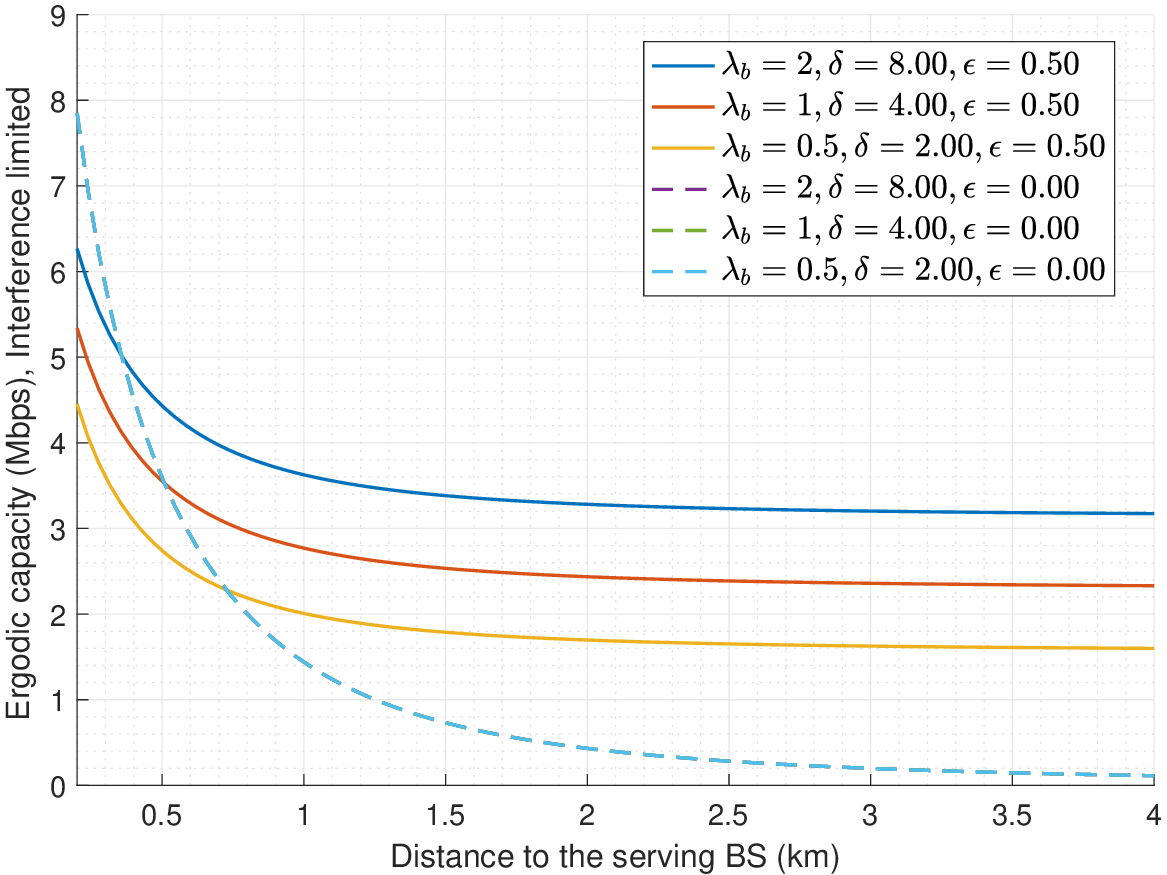} \label{fig:ergCapIL_sameRatio_ratio025_epsilon}}

    \caption{(a)-(f) Ergodic capacity for the noise-limited and the interference-limited systems as a function of the distance $r$ between any randomly selected user and its serving base station. Figure (a) analyzes the effect of the power control coefficient $\epsilon$ while taking into consideration the distribution of $r$. Figure (b) evaluates the ergodic capacity for different number of receiver antennas $M$. Figure (c) analyzes the effect of the density of base stations $\lambda_{\mathrm{b}}$ for a constant reuse factor $\delta$ while taking into consideration the distribution of the maximum distance between cell-edge users and their serving base station $r_{\max}$. Figure (d) evaluates the ergodic capacity for different $\lambda_{\mathrm{b}}$ and $\delta$, with proportional ratios of $\lambda_{\mathrm{b}}/\delta$. Finally, Figures (c) and (d) analyzes the effect of $\epsilon$ for different $\lambda_{\mathrm{b}}$ and $\delta$ while taking into consideration the distribution of $r_{\max}$.}
    \label{fig:simulation_results_partA}
\end{figure}

Figure~\ref{fig:ergCapNL_sameM} shows the ergodic capacity of a noise-limited system as a function of the user–base-station distance $r$ for different fractional power control coefficients $\epsilon$ when $B=1$ MHz and $M=1$. As we can see, the choice of $\epsilon \in [0,1]$ significantly changes the ergodic capacity experienced by the users within the same cell. Small $\epsilon$ favors cell-center users but severely penalizes cell-edge users, whereas high $\epsilon$ equalizes performance up to the threshold $r_{\text{th}} = (\bar{P}/P)^{1/(\alpha\epsilon)}$, beyond which users hit the peak-power limit and the capacity drops. When overlaid with the CDF of user distances (dashed lines), it becomes evident that dense networks ($\lambda_{\mathrm{b}} \uparrow$) benefit from low $\epsilon$, whereas sparse networks require higher $\epsilon$ to support users farther away. The density of base stations is therefore determinant when selecting the optimal value of $\epsilon$. For more information on the selection of $\epsilon$, and why selecting a single network-wide power control coefficient is better than having multiple user-based power control coefficients, one can read~\cite{novlan2013analytical}. For our case scenario, we simply consider $\epsilon = 0.5$, as it provides a robust compromise between the benefits of $\epsilon=0$ and $\epsilon=1$, and results in an acceptable ergodic capacity for the majority of users in the network.

Figure~\ref{fig:ergCapNL_sameEpsilon} shows the ergodic capacity of a noise-limited system as a function of the distance $r$ for different number of receiver antennas $M$ when $B=1$ MHz. As is the case in SIMO systems, users benefit from the spatial multiplexing at the receiver, resulting in higher ergodic capacities for increasing $M$. Specially, notice that we have purposely selected four powers of $4$ for the different values of $M$. This is to show that the ergodic capacity increases logarithmically for $M$, as the multiplexing gain comes from the effective scalar channel gain, which appears inside the logarithm when calculating the ergodic capacity.

Figure~\ref{fig:ergCapIL_diffRatio_delta4} shows the ergodic capacity of an interference-limited system as a function of the distance $r$ for different density of base stations $\lambda_{\mathrm{b}}$ and for different number of receiver antennas $M$ when $B=1$ MHz and $\delta=4$. Notice that higher values of $\lambda_{\mathrm{b}}$ represent denser networks, characterized by smaller coverage areas and shorter distances between users and their serving base stations. Specially, since $\delta$ is not adjusted proportionally to $\lambda_{\mathrm{b}}$, higher ratios of $\lambda_{\mathrm{b}}/\delta$ lead to higher interference and lower ergodic capacities. If we also consider the CDF of the maximum distance between the users and their serving base station, we observe that denser networks exhibit larger capacity disparities among cell-edge users (up to $1.5$~Mbps spread for $\lambda_{\mathrm{b}}=2$~BS/km² compared to less than $1$~Mbps for $\lambda_{\mathrm{b}}=0.5$~BS/km²). This results highlights a fairness--throughput trade-off, where denser networks lead to higher disparities among cell-edge users, whereas sparser networks lead to a more uniform experience among cell-edge users.

Figure~\ref{fig:ergCapIL_sameRatio_ratio025} shows the ergodic capacity of an interference-limited system as a function of distance $r$ for different number of receiving antennas $M$ when the reuse factor scales proportionally with base-station density ($\delta \propto \lambda_{\mathrm{b}}$) and $B = 1$ MHz. Under this adaptive reuse strategy, the ratio $\lambda_{\mathrm{b}}/\delta = 0.25$ is held constant, so increasing $\lambda_{\mathrm{b}}$ yields denser deployments with smaller cells and shorter average user–base-station distances, while the number of co-channel interferers remains unchanged. The results reveal that ergodic capacity grows with $\lambda_{\mathrm{b}}$ and increases further with the number of receive antennas $M$. Notably, a denser network equipped with fewer antennas per base station outperforms a sparser network with more antennas when $\delta$ and $\lambda_{\mathrm{b}}$ are scaled proportionally. Moreover, comparing Figures~\ref{fig:ergCapIL_diffRatio_delta4} and~\ref{fig:ergCapIL_sameRatio_ratio025} reveals that the ergodic capacity is considerably more sensitive to the reuse factor $\delta$ than to base-station density $\lambda_{\mathrm{b}}$ alone, underscoring the pivotal role of spectrum partitioning in interference-limited edge-intelligent networks.

Finally, Figures~\ref{fig:ergCapIL_diffRatio_delta4_epsilon} and~\ref{fig:ergCapIL_sameRatio_ratio025_epsilon} show the ergodic capacity of an interference-limited system as a function of the distance $r$  for varying fractional power control coefficients $\epsilon$ and different combinations of $\lambda_{\mathrm{b}}$ and $\delta$. Notice that we have selected the power control coefficients in the range $\epsilon \in[0,0.5]$, as these are the values for which our interference-limited model applies; other values outside this range would correspond to a system where the received signals are equally dominated by noise and interference. Notice also that systems with the same ratio $\lambda_{\mathrm{b}}/\delta$ and $\epsilon = 0$ are analytically equivalent and exhibit identical capacity curves. Considering both figures, we observe that higher $\epsilon$ provides high ergodic capacities for users in small cells but degrades the ergodic capacities for users farther away. Again, $\epsilon=0.5$ emerges as a compromise for fairness in interference-limited cases without sacrificing average throughput.

\section{End-to-end offloading model}
\label{sec:offloading_model}
Having established the network model in Section~\ref{sec:network_model} and analyzed its dependence on key network parameters, such as base station density $\lambda_{\rm{b}}$, frequency reuse factor $\delta$, power-control coefficient $\epsilon$, and number of receive antennas $M$, we now characterize the end-to-end delay experienced by video-analytics users. As illustrated in Figure 3, the offloading process consists of three interdependent stages reflecting both wireless and computational dynamics: the uplink transmission of each frame, the queueing at the edge server, and the execution of the AI inference task. To enable a tractable analysis of this multi-stage process, we decompose the end-to-end delay as
\begin{equation}
    T_\mathrm{ul} + T_\mathrm{w} + T_\mathrm{s}
\end{equation}
where $T_\mathrm{ul}$ is the time to transmit a frame, $T_\mathrm{w}$ is the time spent in the queue, and $T_\mathrm{s}$ is the time to process the frame. We derive below closed-form expressions for each of these delay components.
\begin{figure}[t]
    \centering
    \resizebox{\columnwidth}{!}{
    \input{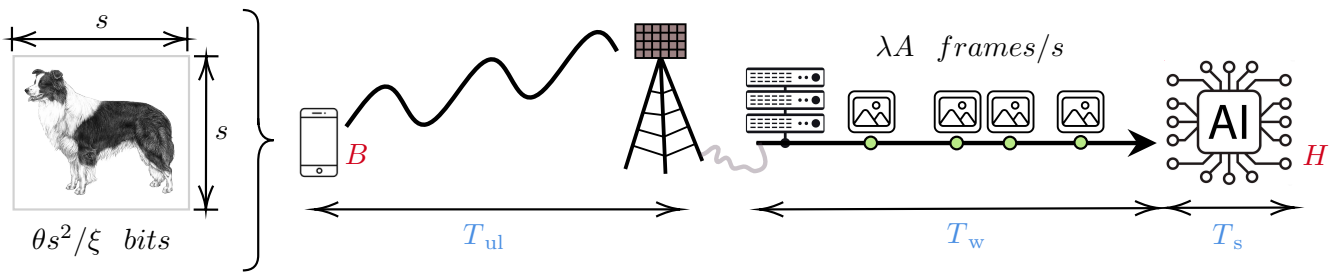}
    }
    \caption{End-to-end offloading timeline for a video-analytics task. Parameters in red correspond to the wireless and computing resources that need to be optimized to satisfy the network, user, and video analytic requirements. Parameters in blue represent (from left to right) the uplink, waiting, and service times of the end-to-end offloading process.}
    \label{fig:basicModel}
\end{figure}

The offloading process begins with the uplink transmission of a compressed video frame. Each user transmits frames of resolution $s \times s$ pixels, originally encoded at $\theta$ bits per pixel and compressed by a factor $\xi:1$. This yields a payload of $\theta s^2 / \xi$ bits per frame. Therefore, for a user located at distance $r$ from its serving base station and allocated bandwidth $B$, the uplink transmission time is
\begin{equation}
    T_{\mathrm{ul}} = \frac{\theta s^2}{\xi \, \phi(B,r)},
    \label{eq:uplink_time}
\end{equation}
where $\phi(B,r)$ is the achievable transmission rate. Depending on the operating regime, $\phi(B,r)$ is given by either of the ergodic capacities derived in Section~\ref{sec:system-model_ULtransmission}:
\begin{equation}
    \phi(B,r) =
    \begin{cases}
        C_{\mathrm{NL}}(B,r) & \text{Noise-limited, defined in \eqref{eq:ergodicCapacity_noiseLimited}}, \\
        C_{\mathrm{IL}}(B,r) & \text{Interference-limited, defined in \eqref{eq:ergodicCapacity_interferenceLimited}}.
    \end{cases}
    \label{eq:uplink_time_capacity}
\end{equation}

Having derived the uplink transmission delay $T_{\rm ul}$, we now characterize the arrival process at the edge server, which depends on the spatial distribution of these delays. The frame generation by user devices is modeled as a spatio-temporal Poisson point proces with intensity $\lambda\, \left[\text{frames}\cdot \mathrm{s}^{-1}\cdot \mathrm{km}^{-2}\right]$. This assumes that frame generation events occur independently across space and time, a standard abstraction in large-scale wireless networks with uncoordinated users~\cite{andrews2011tractable}. Since users are uniformly distributed within Voronoi cells and transmit independently, the uplink transmission delays $T_{\rm ul}$ are i.i.d.\ conditional on user locations. By the \emph{displacement theorem} for Poisson point processes~\cite[Theorem~1.3.9]{baccelli2009stochastic}, independent random displacements in time preserve the Poisson structure of the arrival process. Consequently, for an edge server whose Voronoi region has random area $A$, the aggregate arrival process is Poisson distributed with intensity $\lambda A$.

%Note that, while real deployments may exhibit some degree of burstiness, the Poisson assumption remains robust in this setting for two reasons. First, each edge server aggregates the traffic of many independent users, and, by the \textit{Palm–Khintchine theorem}, the superposition of many independent arrival streams converges to a Poisson process even when each individual stream exhibits mild correlation or burstiness~\cite{heiman1984stochastic}. Second, the Poisson process maximizes the entropy of inter-arrival times for a given mean rate, yielding a conservative (worst-case) bound on delay variance~\cite{kleinrock1976queueing}. Hence, the use of a Poisson arrival model is both analytically tractable and pessimistic in terms of delay guarantees.

Once a frame is fully received at the base station, it is forwarded to the co-located edge server over a high-speed internal link, whose propagation delay is negligible relative to the uplink transmission and inference times. Upon arrival at the server, each frame is placed in an infinite buffer and processed according to a first-come–first-served basis. The processing time $T_\mathrm{s}$ is deterministic and is determined by the resolution of the transmitted frames and the available computational resources at the edge server. In particular, to perform the video-analytics inference on the offloaded frames, the edge server utilizes a YOLOv5-based convolutional neural network~\cite{glenn2021yolov5}, selected for its efficiency in real-time object detection on resource-constrained edge servers. Since YOLOv5 is capable of handling frames of different resolutions without changing its associated learnable parameters, it allows parameterization of the inference time. Following well-established profiling results for YOLOv5-based inference~\cite{liu2018edge}, the service time is
\begin{equation}
    T_\mathrm{s} = \frac{c_1s^3 + c_2}{H},
    \label{eq:service_time}
\end{equation}
where $H$ is the available computational processing frequency at the server (in TFLOPS/s), and $c_1$, $c_2$ are positive architecture-specific constants. The cubic dependence arises because the dominant operations (convolutions on feature maps) scale nearly cubically with the input frame resolution. This model is widely used in performance evaluation of edge-based deep learning and has been validated on both GPU- and accelerator-based edge servers~\cite{wang2022performance}.

Similarly, it is possible to parameterize the accuracy of the object detection algorithm as
\begin{equation}
    a(s) = c_3 - c_4e^{-c_5 s},
    \label{eq:accuracy_YOLO}
\end{equation}
for some other architecture-specific positive constants $c_3$, $c_4$, and $c_5$, where the accuracy of the detection is measured as the mean average precision of the object detection algorithm for a predefined threshold of the intersection over union~\cite{liu2018edge}. For more information on the parameterization of the deep learing algorithm and the effect of the learning and inference processes on the constants $c_1,\dots,c_5$, refer to~\cite{bochkovskiy2020yolov4}.

The combination of Poisson arrivals and deterministic service yields a M/D/1 queueing system with load $\rho = \lambda A\, T_\mathrm{s} < 1$ required for stability. Considering this M/D/1 queue, the complementary cumulative distribution function (CCDF) of the waiting time $T_\mathrm{w}$ can be derived from the state probabilities using the Erlang’s principle of statistical equilibrium~\cite[Sections 10.4.2 and 10.4.4]{iversen2010teletraffic}. Specially, for a given load $\rho$ and service time $T_\mathrm{s}$, the CCDF of the waiting time is
\begin{equation}
    \mathcal{P}(T_\mathrm{w}>T)=1-(1-\rho)\sum_{\nu=0}^{\Tilde{T}}\frac{\left[\rho(\nu-t)\right]^\nu}{\nu!}\,e^{-\rho(\nu-t)},
    \label{eq:CCDF_wait}
\end{equation}
for any $T\geq 0$, where $t = \frac{T}{T_\mathrm{s}}$, and $\Tilde{T}=\left\lfloor t \right\rfloor$ is the greatest integer that is less than or equal to $t$. This function is continuous and smooth, increasing monotonically for increasing $\rho$ (higher loads lead to longer queues), and decreasing monotonically for increasing $T_\mathrm{s}$ (faster service reduces queue backlog). Notably, it is concave in $T\in [0,T_\mathrm{s})$ and convex in $T\in [T_\mathrm{s},\infty)$. These curvature transitions are critical when embedding~\eqref{eq:CCDF_wait} into the resource-dimensioning problem, as they create non-convex feasible sets that must be handled with care.

Overall, these closed-form expressions for $T_{\rm ul}$, $T_{\rm w}$, and $T_{\rm s}$ provide a tractable yet realistic framework for system optimization. In the next section, we leverage this model to formulate and solve a joint resource-dimensioning problem, minimizing deployed compute $H$ and bandwidth $B$ resources while satisfying probabilistic delay and minimum accuracy constraints under spatial randomness.

\begin{remark}
    Even though \eqref{eq:CCDF_wait} is derived under Poisson arrivals (yielding an M/D/1 queue), real deployments may exhibit some degree of burstiness due to protocol aggregation, scheduler coupling, or correlated sensing. A standard way to capture such temporal correlation is to model the arrival stream by a Markovian Arrival Process, resulting instead in an MAP/D/1 queue.  Importantly, the qualitative curvature transition of \eqref{eq:CCDF_wait} around the deterministic service time $T_s$ is induced by deterministic service rather than by the Poisson assumption itself. Accordingly, under MAP/D/1 the waiting-time distribution retains an analogous $T_s$-induced piece-wise  structure, with burstiness primarily affecting the magnitude of $\mathcal{P}(T_{\mathrm{w}}>T)$ for a given load. Consequently, the M/D/1 queue serves as a representative baseline that yields a tractable closed form and makes explicit the $T_s$-driven behavior, and burstiness can be incorporated by replacing \eqref{eq:CCDF_wait} with the corresponding MAP/D/1 waiting-time tail without changing the fundamental structure of the offloading framework.
\end{remark}

\section{Resource-dimensioning problem}
\label{sec:resource_dimensioning}
In this section, we address the resource-dimensioning problem for the considered multi-cell edge video-analytic system. Our goal is to determine the optimal bandwidth $B$ allocated per frame transmission and the optimal computing capacity $H$ allocated at each edge-intelligent server, such that the system minimizes the costs per unit area while satisfying statistical quality-of-service requirements. These requirements serve as key benchmarks for 5G/6G networks and encompass latency, coverage for cell-edge users and servers, and accuracy of the video analytics. This focus is essential because inadequate dimensioning of resources in edge computing systems can result in excessive operational costs, degraded performance for delay-sensitive applications, and reduced energy efficiency~\cite{premsankar2022energy}. Considering this, we first outline the quality-of-service requirements that guide our approach, followed by the formulation of the optimization problem, and conclude with a theoretical analysis of its properties.

\subsection{Quality-of-service requirements}
In the considered edge-analytic systems, all frames receive the same bandwidth resources, all servers are equipped with the same computing resources, and all users use the same power control coefficient. Besides, all users can transmit their frames, even those who are very far away from their base station and could know that their associated server would process their frames late. As a result, the users that experience the largest $T_\mathrm{ul}$ are the ones located at the cell edge, and the users that experience the largest $T_\mathrm{w}$ and $T_\mathrm{s}$ are the ones located in large cell areas. Notice, however, that these types of users are not necessarily mutually exclusive, as the Voronoi cells adopt irregular shapes, with users being far from the base station, but nonetheless within large or small cell areas. Therefore, the optimization problem is formulated under the constraints that the frames from the users at the cell edge and in large Voronoi cells satisfy the quality requirements. That means, if the quality requirements of these users are satisfied for some $B$ and $H$, then the quality requirements of all the other users at closer distances to the base station and/or in smaller cell areas will be satisfied as well for the same $B$ and $H$. This worst-case approach ensures comprehensive coverage and QoS across the network, preventing scenarios where central users benefit at the expense of edge users, which could otherwise lead to unfair service distribution and potential system instability under high loads.

As for the entire end-to-end offloading process, we define the following:

\emph{\textbf{Latency requirement}}: The delay of the entire offloading process, which starts when a user begins offloading a frame and ends when the server finishes processing that frame, should be less than the maximum delay requirement $D$ with a minimum probability $\omega_\mathrm{min}$ for all frames of all covered users. Formally,
\begin{equation}
    \mathcal{P}\left(T_\mathrm{ul}+T_\mathrm{w}+T_\mathrm{s}\leq D\right) \, \geq \,\omega_{\min}.
    \label{eq:QR1-probSuccessfulComputation}
\end{equation}
From the analysis in Section~\ref{sec:offloading_model}, it follows that this probability can be calculated from the CCDF of the waiting time, and the mathematical expression for uplink transmission time and service time, defined in \eqref{eq:uplink_time} and \eqref{eq:service_time}, respectively. For simplicity, we consider that all users in the network have the same delay requirements $D$ and $\omega_{\min}$ for all frames and the same payload for all image processing tasks. Other cases are possible, but allowing for different delay requirements or payloads in the system requires a more intricate network and server model, a topic we recognize as a potential future research.

To ensure coverage,

\emph{\textbf{QoS cell-edge users}}: The optimal bandwidth resources should ensure that at least $\eta_r$ of the users located the furthest from their serving base station have enough wireless resources to satisfy the latency requirement. Focusing on cell-edge users is justified because they experience the highest path losses and lowest signal strengths, making them the bottleneck for network performance. Thus, addressing their needs guarantees adequate service for all users. Formally, we consider that the optimal bandwidth should be calculated considering that the maximum distance between any user and its serving base station, $r_{\max}$, satisfies
\begin{equation}
    \mathcal{P}\left(r_{\max}\leq \sqrt{\lambda_{\rm b}}\,x \;\big| \;\alpha, \beta, \gamma\right) \geq \eta_r.
    \label{eq:QR2-maxRadiusConstraint}
\end{equation}

\emph{\textbf{QoS servers}}: The optimal computing resources should ensure that at least $\eta_A$ of the servers in the system have enough computational resources to process the incoming traffic from its associated users and satisfy the latency requirements. This server-centric QoS is crucial in multi-cell scenarios where cell sizes vary due to the PPP model, leading to uneven load distribution. Ensuring a high percentage of servers meet the requirements prevents overload in densely populated areas and maintains overall system reliability. Formally, we consider that the optimal computing capacity should be calculated considering that the cell area from which an edge server processes its incoming traffic, $A$, satisfies
\begin{equation}
    \mathcal{P}\left(A \leq \lambda_{\rm b}\,x \;\big| \;\alpha, \beta, \gamma\right) \geq \eta_A.
    \label{eq:QR3-maxAreaConstraint}
\end{equation}

Finally, considering the inference of the video analytics,

\emph{\textbf{Accuracy requirement}}: The accuracy of the AI-based object detection algorithm should be above a minimum threshold $a_\mathrm{min}$ for all frames of all covered users. Maintaining high accuracy is imperative because suboptimal resource allocation could force compromises in image quality, directly impacting the utility of edge intelligence applications. Formally, we consider from the expression in \eqref{eq:accuracy_YOLO} that the resolution of the transmitted frames should satisfy
\begin{equation}
    s \geq \frac{1}{c_5}\ln\left(\frac{c_4}{c_3-a_{\min}}\right).
    \label{eq:QR4-minAccuracyConstraint}
\end{equation}

Altogether, these latency, QoS, and accuracy requirements affect the entire offloading process, and they are crucial to determining the optimal wireless and computing resources. By incorporating these constraints, our formulation provides a balanced and justifiable framework that aligns with practical deployment needs and mitigates risks of under-provisioning or over-provisioning wireless and computational resources.

\subsection{Optimization problem}
After formalizing the QoS requirements, we are now ready to formulate the resource-dimensioning problem. Our objective is to minimize the total cost of wireless and computing resources per unit area in the multi-cell edge video-analytic system. To achieve this, we adopt the weighted sum method~\cite{marler2010weighted} via the trade-off parameter $\beta_1 \in [0,1]$ to balance the relative costs of bandwidth and computing capacity, assuming normalized units via a positive parameter $\beta_2$. This approach is justified as it allows for flexible weighting based on deployment-specific cost factors, such as spectrum licensing fees versus hardware expenses, ensuring the model's applicability across diverse scenarios without loss of generality.

The problem is formulated as a joint optimization over the bandwidth $B$, computing capacity $H$, auxiliary variable $T$, distance to the serving base station $r$, cell area $A$, and frame width/length $s$ for a fixed arrival rate of tasks $\lambda$, base station density $\lambda_{\rm b}$, frequency reuse factor $\delta$, delay requirement $D$, minimum success probability $\omega_{\min}$, coverage probabilities $\eta_r$ and $\eta_A$, and minimum accuracy $a_{\min}$. Specifically, the optimization problem is given by
\begin{subequations}
    \begin{align}
    \underset{\{B, H, T, r, A, s\}}{\textrm{minimize}} \quad & \beta_1\lambda B + (1-\beta_1)\beta_2\lambda_{\mathrm{b}} H + \vartheta  A \label{eq:objective_function} \\
    \textrm{subject to} \;\;\quad & \mathcal{P}(T_\mathrm{w} > T) \leq 1 - \omega_{\min}, \label{eq:constrain1}\\
                        & \frac{s^2 \kappa_1}{\phi(B,r)} + T + \frac{c_1 s^3 + c_2}{H} = D, \label{eq:constrain2}\\
                        & A\, (c_1 s^3+c_2) \kappa_2 \leq H, \label{eq:constrain3}\\
                        & B > 0, \; H > 0, \; T \geq 0,  \label{eq:constrain4}\\
                        & r \geq \kappa_3, \; A \geq \kappa_4,\; s \geq \kappa_5, \label{eq:constrain5}
    \end{align}
    \label{eq:optimization_problem_original}
    %& \frac{\kappa_1}{B \exp\left(B\kappa_2\,r^{\alpha}\right) E_1\left(B\kappa_2\,r^{\alpha}\right)} + T +  \frac{\kappa_3}{H} \leq D, \label{eq:constrain2}\\
    %\vspace{-.2cm}
\end{subequations}
where $\phi(B,r)$ is defined in~\eqref{eq:uplink_time_capacity}, and 
\begin{equation}
    \kappa_1 = \frac{\theta}{\xi},\quad \kappa_2 = \frac{\lambda}{\rho_{\max}}, \quad \kappa_3 = \frac{\mathcal{P}_{\bar{r}_{\max}}^{-1}(\eta_r)}{\sqrt{\lambda_{\mathrm{b}}}}, \quad \kappa_4 = \frac{\mathcal{P}_{\bar{A}}^{-1}(\eta_A)}{\lambda_{\mathrm{b}}},\quad \kappa_5 = \frac{1}{c_5}\ln\left(\frac{c_4}{c_3-a_{\min}}\right),
    \label{eq:constants}
\end{equation}
are constants that depend on the network parameters defined in Section~\ref{sec:network_model}.

Constraint \eqref{eq:constrain1} ensures the queuing delay meets the probabilistic latency requirement, derived from the M/D/1 queue analysis in Section~\ref{sec:offloading_model}. Constraint \eqref{eq:constrain2} enforces the deterministic end-to-end delay bound, incorporating the uplink transmission time $T_\mathrm{ul}$ (which differs for noise-limited and interference-limited cases as per ~\eqref{eq:uplink_time}) and service time $T_\mathrm{s}$ \eqref{eq:service_time}. Constraint \eqref{eq:constrain3} guarantees that the server satisfies the demands of all users and does not overload. Constraint \eqref{eq:constrain4} limits the domain of the optimization variables. Finally, constraint \eqref{eq:constrain5} guarantees coverage for cell-edge users and servers using the inverse CDFs from stochastic geometry models in \eqref{eq:QR2-maxRadiusConstraint} and \eqref{eq:QR3-maxAreaConstraint}, and secures the minimum accuracy via the resolution-accuracy relationship in \eqref{eq:QR4-minAccuracyConstraint}.

This formulation is non-convex because constraints \eqref{eq:constrain1}, \eqref{eq:constrain2}, and \eqref{eq:constrain3} are non-convex. To tackle this efficiently, we resort to breaking down the resource-dimensioning problem into a series of sub-problems, each containing only a small set of optimization variables and constraints. Then, by solving these sub-problems sequentially, we aim to establish feasibility and optimality guarantees for the original problem.

\subsection{Theoretical analysis}
\label{subsec:theoretical_analysis}
We now delve into a more detailed theoretical analysis of the non-convex constraints of the optimization problem to establish its structural properties and derive conditions under which the global optimum can be efficiently obtained through convex reformulation.

First, for the waiting time constraint \eqref{eq:constrain1}, recall from the analysis in Section~\ref{sec:offloading_model} that the CCDF of the waiting time is monotonically increasing for increasing $\rho$, monotonically decreasing for increasing $T_\mathrm{s}$, concave in $T\in [0,T_\mathrm{s})$, and convex in $T\in [T_\mathrm{s},\infty)$. Besides, if we combine the mathematical expression of the server load, $\rho = \lambda A\, T_\mathrm{s}$, with the definition of the service time \eqref{eq:service_time},
\begin{equation}
    \rho = \lambda A\, \frac{c_1s^3 + c_2}{H},
    \label{eq:loadServer_expanded}
\end{equation}
we can further conclude that the CCDF of the waiting time is monotonically increasing for increasing $A$ and $s$, and monotonically decreasing for increasing $H$.

For the end-to-end delay constraint \eqref{eq:constrain2}, the ergodic capacities in both the noise-limited and interference-limited regimes (defined in \eqref{eq:ergodicCapacity_noiseLimited} and \eqref{eq:ergodicCapacity_interferenceLimited}, respectively) are concave in $B$. Considering this, let us define the following functions,
\begin{equation*}
    f(x) = h(g(x)),\quad h(x) = \frac{1}{x}, \quad g(x) = \phi(x,r),\quad x>0,
\end{equation*}
where $f(x)$ represents the multiplicative inverse function of the ergodic capacity in terms of $B$, and $g(x)$ represents the ergodic capacity of the system for a given distance $r$. Since $h(x)$ is convex and non-increasing, and $g(x)$ is concave in $B$, it follows from~\cite[Section~3.2.4]{boyd2004convex} that $f(B)$ is convex in $B$. Note also that the ergodic capacities are convex in $r$. Considering this, let us define another set of functions,
\begin{equation*}
    \Tilde{f}(x) = \Tilde{h}(\Tilde{g}(x)), \quad \Tilde{h}(x) = - \frac{1}{x}, \quad \Tilde{g}(x) = -\phi(B,x),\quad x>0
\end{equation*}
where $\Tilde{f}(x)$ represents the multiplicative inverse function of the ergodic capacity in terms of $r$, and $\Tilde{g}(x)$ represents the negative ergodic capacity of the system for a given bandwidth $B$. Since $\Tilde{h}(x)$ is concave and non-decreasing, and $\Tilde{g}(x)$ is concave in $r$, it follows from~\cite[Section~3.2.4]{boyd2004convex} that $\Tilde{f}(r)$ is concave in $r$. From this analysis, we can conclude that constraint \eqref{eq:constrain2} is convex in $B$, $T$, and $H$, and concave in $r$. Besides, if we consider the terms in the numerators of the first and third summands, we can further conclude that constraint  \eqref{eq:constrain2} is convex in $s$.

For constraint \eqref{eq:constrain2}, note that the non-convexity comes from the multiplication of the two optimization variables $A$ and $s$. Otherwise, for a fixed $s$, the constraint is convex.

The above monotonicity properties imply that the CCDF of the waiting time \eqref{eq:constrain1} and the delay of the entire offloading process \eqref{eq:constrain2} are monotonically increasing for increasing $r$ and $s$. Conversely, both of these constraints are monotonically decreasing for increasing $B$ and $H$. Since we are minimizing the overall network resources, that optimal solutions for $r$ and $s$ are given by their corresponding boundaries, defined in \eqref{eq:constrain5}. We therefore fix $r = r_{\max} = \kappa_3$ and $s = \kappa_5$, where $\kappa_3$ and $\kappa_5$ are constants obtained by inverting the corresponding CDFs and the accuracy-log-resolution relationship.

With $r$ and $s$ fixed, the only remaining source of non-convexity is the equality in \eqref{eq:constrain2} and the piece-wise definition of the power control scheme in $\ell(r, \alpha, \epsilon)$. We address the equality in \eqref{eq:constrain2} by relaxing it with an inequality, and address the pice-wise dependence of the ergodic capacity on the peak power constraint \eqref{eq:peakPowerConstrain} by utilizing the epigraph representation of the $\min$ operator. For the latter, we specifically introduce two separate delay constraints: one for the fractional power control regime ($\phi_{\textit{low}}$) and one for the peak-power capped regime ($\phi_{\textit{peak}}$), both of which are convex in the optimization variables as the epigraph is an operation that preserve convexity~\cite[Section~3.2.3]{boyd2004convex}.

Putting all together, the reformulated optimization problem can be expressed over the bandwidth $B$, computing capacity $H$, auxiliary variable $T$, and cell area $A$ for a fixed arrival rate of tasks $\lambda$, base station density $\lambda_{\rm b}$, frequency reuse factor $\delta$, delay requirement $D$, minimum success probability $\omega_{\min}$, coverage probabilities $\eta_r$ and $\eta_A$, and minimum accuracy $a_{\min}$ as

\begin{subequations}
    \begin{align}
    \underset{\{B,H,T,A\}}{\textrm{minimize}} \quad & \beta_1\lambda B + (1-\beta_1) \beta_2 \lambda_{\mathrm{b}} H + \vartheta  A \label{eq:objective_function_convex} \\
    \textrm{subject to} \;\;\quad & \mathcal{P}(T_\mathrm{w} > T) \leq 1 - \omega_{\min}, \label{eq:constrain1_convex}\\
                        & \frac{\kappa_5^2\, \kappa_1}{\phi_{\textit{low}}(B,\kappa_3)} + T + \frac{c_1 \kappa_5^3 + c_2}{H} \leq D, \label{eq:constrain2_convex}\\
                        & \frac{\kappa_5^2\, \kappa_1}{\phi_{\textit{peak}}(B,\kappa_3)} + T + \frac{c_1 \kappa_5^3 + c_2}{H} \leq D, \label{eq:constrain3_convex}\\
                        & A\, (c_1 \kappa_5^3+c_2) \kappa_2 \leq H, \label{eq:constrain4_convex}\\
                        & B > 0, \; H > 0, \; T \geq 0,\; A \geq \kappa_4, \label{eq:constrain5_convex}
    \end{align}
    \label{eq:optimization_problem_convex}
\end{subequations}
where
\begin{align*}
    \phi_{\textit{low}}(B,\kappa_3) &= \phi(B,\kappa_3) \quad \text{when} \quad \ell(r,\alpha,\epsilon) = Pr^{\alpha\epsilon}, \\
    \phi_{\textit{peak}}(B,\kappa_3) &= \phi(B,\kappa_3) \quad \text{when} \quad \ell(r,\alpha,\epsilon) = \Bar{P},
\end{align*}
are the ergodic capacities in the fractional and peak-power regimes evaluated at $r = \kappa_3$.

\begin{lemma}
    \label{lemma:optimalAnalysis}
    \textit{The solution to the reformulated convex problem \eqref{eq:optimization_problem_convex} is globally optimal for the original non-convex resource-dimensioning problem~\eqref{eq:optimization_problem_original} if at least one of the following conditions holds:
    \begin{equation}
            T^* \geq \frac{c_1\kappa_5^3+c_2}{H^*} \quad \textit{or} \quad \rho^* \geq
            1+W\left(-\frac{\omega_{\min}}{e}\right),
            \label{eq:global_optimality_conditions}
        \end{equation}
        where $e=2.718281\dots$ is the Euler's number, and $W(\cdot)$ is the principal branch of the Lambert function. % \rho^* = \lambda A^* \frac{c_1\kappa_5^3+c_2}{H^*}
    }
    \begin{proof}
        The only non-convexity in \eqref{eq:optimization_problem_convex} is the concavity of the CCDF for $T \in [0, T_\mathrm{s})$. If the optimal solution satisfies $T^* \geq T_\mathrm{s}^*$, the CCDF of the waiting time is convex in the relevant region, and thus the solution is globally optimal. Here $T^*$ can be derived from the definition of the service time \eqref{eq:service_time} and the optimal solution to the frame width/height, $\kappa_5$, given in \eqref{eq:constants}. Alternatively, note that,
        \begin{equation*}
            \max_{T\geq T_\mathrm{s}} \, \mathcal{P}(T_\mathrm{w} > T) = \mathcal{P}(T_\mathrm{w} > T_\mathrm{s}) = 1 - (1-\rho)e^{\rho}.
        \end{equation*}
        Thus, whenever $1-\omega_{\min} \leq 1-(1-\rho)e^\rho$, the feasible set of the waiting-time constraint lies entirely in the convex region $T \geq T_\mathrm{s}$. Solving the inequality for $\rho$ yields the second condition in \eqref{eq:global_optimality_conditions}. In both cases, the non-convex portion of the feasible set is inactive, and the solution to \eqref{eq:optimization_problem_convex} coincides with the global optimum of the original problem. This concludes the proof.
    \end{proof}
\end{lemma}

\begin{remark}
\label{remark:pareto_frontier}
Whenever the global-optimality conditions of Lemma~\ref{lemma:optimalAnalysis} are satisfied, varying the trade-off parameter $\beta_1 \in (0,1)$ in the objective function \eqref{eq:objective_function_convex} generates the complete Pareto frontier of communication cost per unit area $\lambda_b B$ versus computation cost per unit area $\lambda_b H$ under the original statistical QoS constraints. 
%Moreover, because of the spatial homogeneity of the offloading model, the expected number of generated frames per cell is $\lambda A$, and every frame is allocated the same bandwidth $B$ and processed by a server of capacity $H$. Consequently, the per-frame communication cost is exactly $B$ and the per-frame computation cost is exactly $H/(\lambda A)$. Hence, under the global-optimality conditions of Lemma~\ref{lemma:optimalAnalysis}, varying $\beta_1 \in (0,1)$ also generates the complete Pareto frontier of per-frame communication versus per-frame computation resources required to support edge video analytics under the specified statistical QoS constraints.
\end{remark}

\section{Numerical results}
\label{sec:numerical_results}
This section evaluates the resource-dimensioning problem for the multi-cellular SIMO system formulated in Section~\ref{sec:resource_dimensioning}. To provide a comprehensive assessment, we decouple the analysis into the noise-limited and interference-limited regimes and examine how key network parameters influence the design of cost-efficient strategies. The optimization problem \eqref{eq:optimization_problem_convex} is solved using the optimization solver from the SciPy library in Python. Regarding the generalized exponential integral $E_i(\cdot)$ in the noise-limited ergodic capacity, we utilize the rational expressions and recurrence relations provided in~\cite[Eq.~5.1.11, Eq.~5.1.14]{abramowitz1988handbook}.

Unless stated otherwise, all results adopt the system parameters in Table~\ref{tab:system_parameters_simulation}, which reflect the standard technical specifications for a typical 5G urban/suburban~\cite{3gpp38901} and a YOLOv5-based edge video-analytics application~\cite{liu2018edge}. The computation-accuracy trade-off \eqref{eq:accuracy_YOLO} and service-time model \eqref{eq:service_time} adopt the profiling methodology established in~\cite{liu2018edge}, parameterized by $c_1 = 7\!\cdot\!10^{-10}$, $c_2 = 0.083$, $c_3 = 1$, $c_4 = 1.578$, and $c_5 = 6.5\!\cdot\!10^{-3}$. These constants yield a root-mean-square fitting error below $0.03$ for both inference latency and accuracy.

%In all evaluated scenarios, we present the solutions to \eqref{eq:optimization_problem_convex} alongside their corresponding optimal server loads to easily corroborate that the second condition of Lemma~\ref{lemma:optimalAnalysis} is consistently satisfied. Consequently, the derived solutions are globally optimal with zero optimality gap, as rigorously guaranteed by the theoretical analysis in Section~\ref{subsec:theoretical_analysis}.
\begin{table}[t]
    \centering
    \caption{Baseline parameter values used in the numerical evaluations.}
    \begin{tabular}{|l|c|}
        \hline
        Path-loss exponent $\alpha$ & 4 \\ \hline
        Power control coefficient $\epsilon$ & 0.5 \\ \hline
        Peak transmission power $\Bar{P}$ & 23 dBm (200 mW) \\ \hline
        Transmission power per unit distance $P$ & 10 dBm/km (10 mW/km) \\ \hline
        Noise power spectral density $N_0$ & -174 dBm/Hz  \\ \hline
        Carrier frequency $f_c$ & 2.4 GHz  \\ \hline
        Frame encoding rate $\theta$ & 24 bits/pixel \\ \hline
        Frame compression rate $\xi$ & 2 \\ \hline
        Number of receiver antennas $M$ & 16 \\ \hline
        Maximum load at the server $\rho_{\max}$ & $0.99$ Erlangs \\ \hline
        Maximum delay requirement $D$ & $500$ ms\\ \hline
        Minimum probability of success $\omega_{\min}$ & $0.8$ \\ \hline
        Ratio of successful cell-edge users $\eta_r$ & $0.999$  \\ \hline
        Ratio of successful cell areas $\eta_A$ & $0.999$ \\ \hline
        Minimum object detection accuracy $a_{\min}$ & $0.9$ \\ \hline
        Trade-off parameter objective function $\beta_1$ & $0.5$ \\ \hline
        Scaling parameter objective function $\beta_2$ & $10^{-6}$  \\ \hline
        Regularization parameter objective function $\vartheta$ & 1 \\ \hline
    \end{tabular}
    \label{tab:system_parameters_simulation}
\end{table}

\subsection{Noise-limited system}
\begin{figure}[!t]
    \centering
    \subfigure[]{\includegraphics[width=0.44\textwidth]{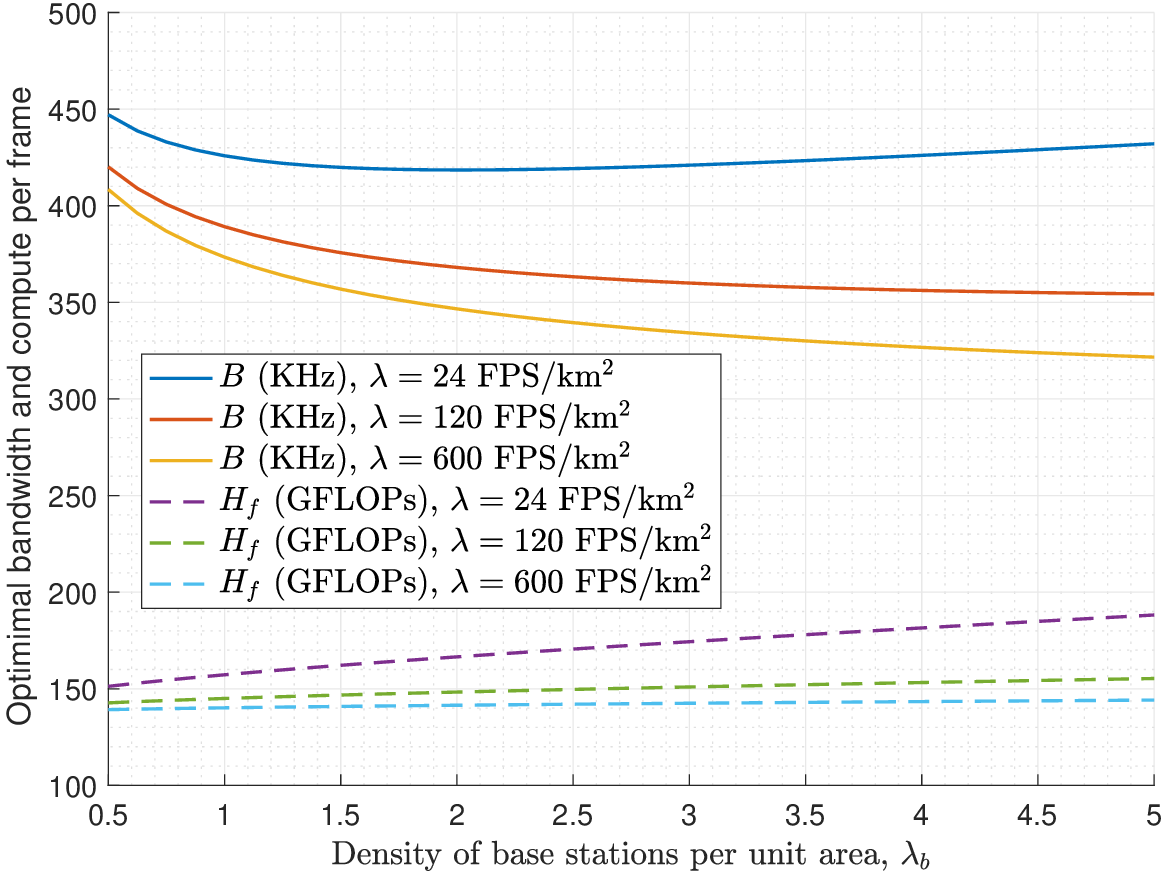} \label{fig:NL_optBH_lambdaBS_lambda}}$\quad$
    \subfigure[]{\includegraphics[width=0.44\textwidth]{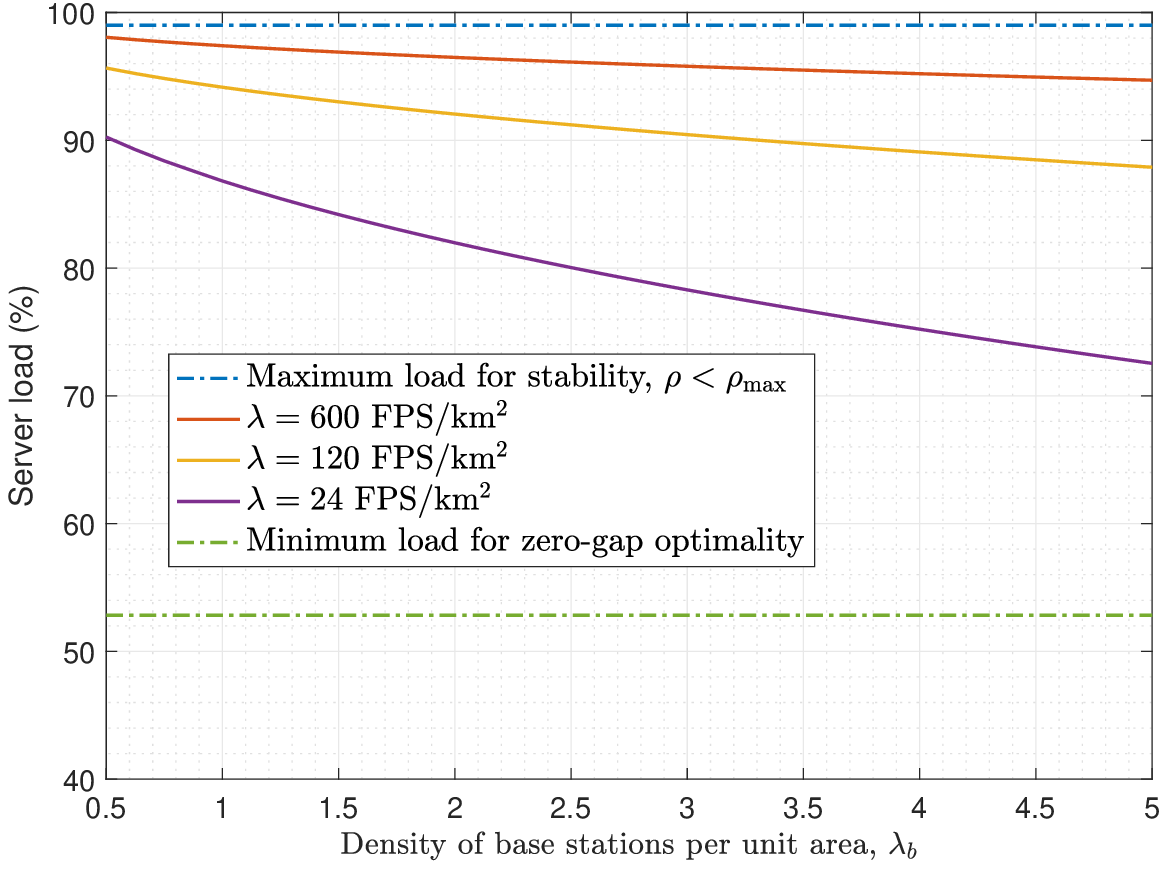} \label{fig:NL_optLoad_lambdaBS_lambda}}
    \subfigure[]{\includegraphics[width=0.44\textwidth]{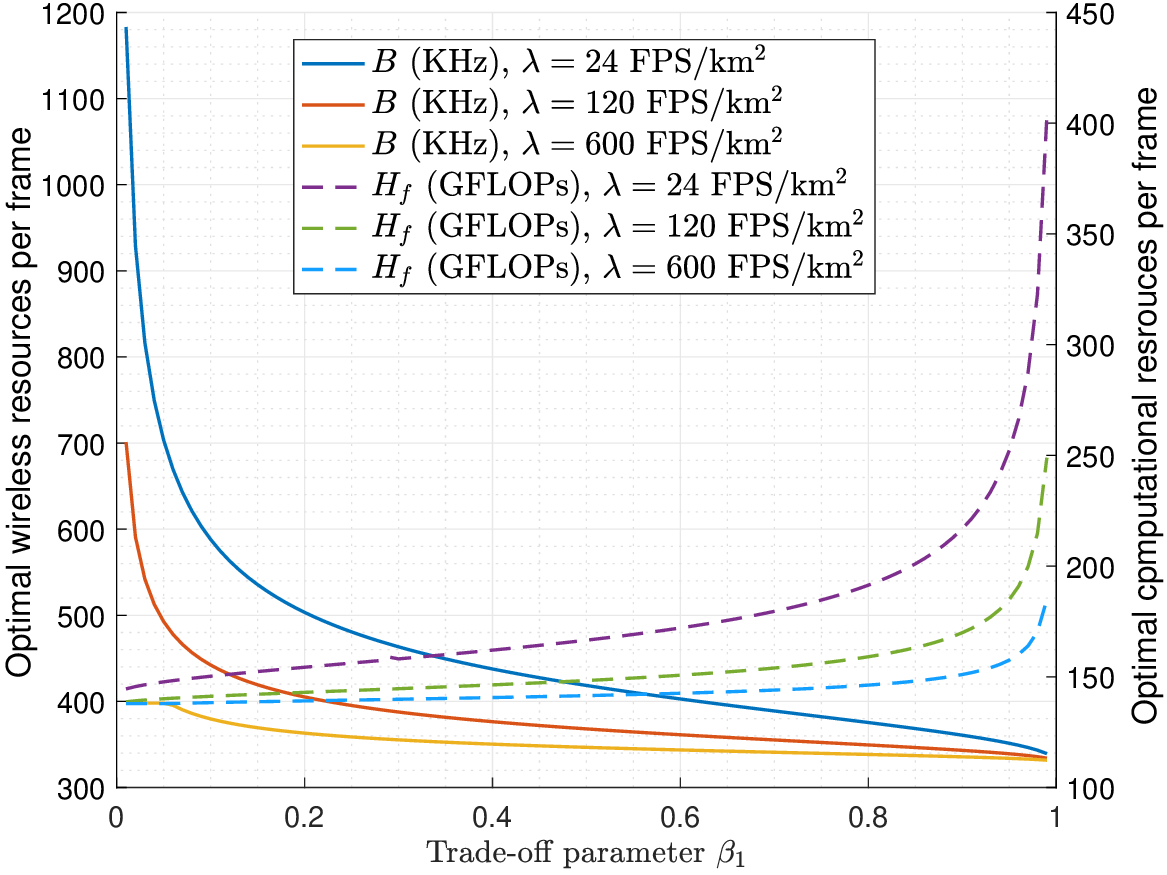} \label{fig:NL_optBH_betaTradeOff}}$\quad$
    \subfigure[]{\includegraphics[width=0.44\textwidth]{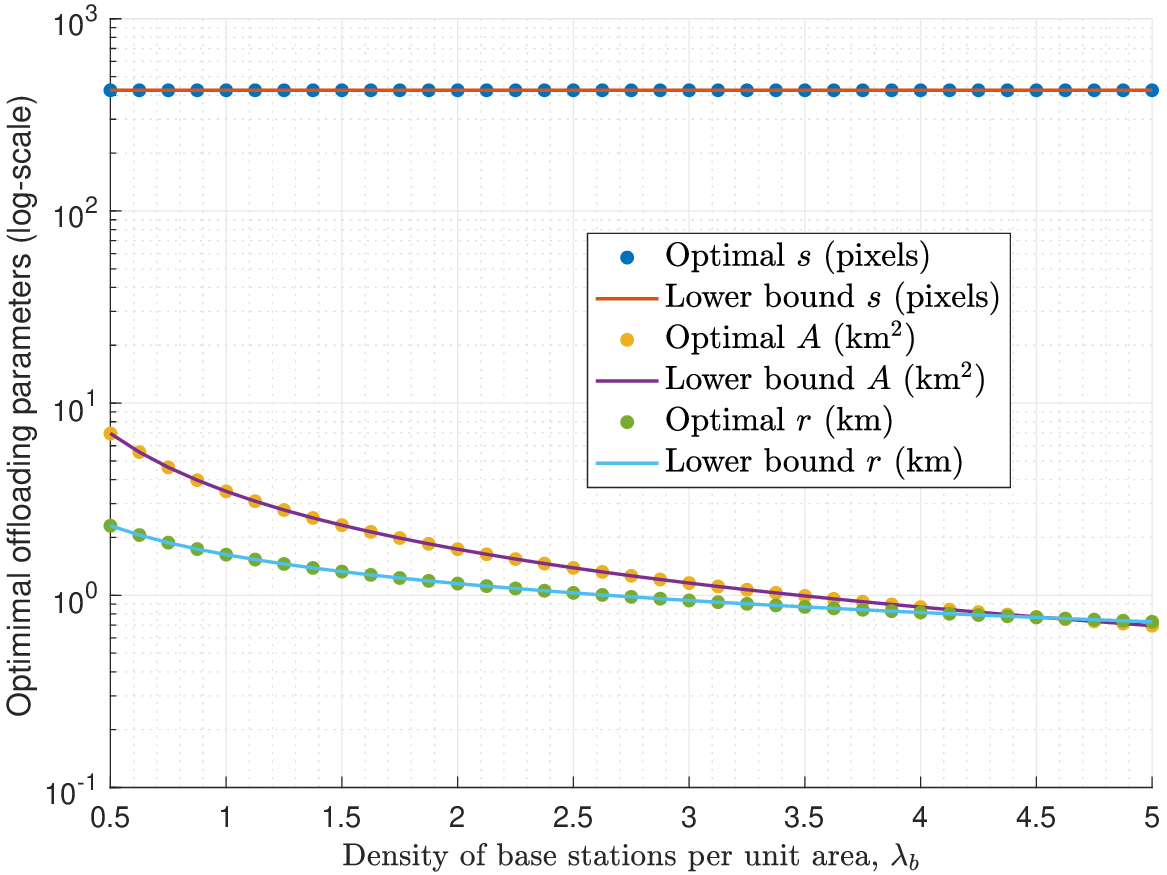} \label{fig:NL_optrAs_lambdaBS}}
    \caption{(a)--(b) Optimal wireless and computing resources per frame, optimal server load, and total cost of the resource-dimensioning (in millions) per unit area for a noise-limited system as a function of the density of base stations per unit area $\lambda_{\mathrm{b}}$. (c) Effect of the trade-off parameter $\beta_1$ on the optimal resources for a noise-limited system with density $\lambda_{\mathrm{b}}=2$ BS/km$^2$. (d) Optimal offloading parameters $r$, $A$, and $s$ for which the optimal $B$ and $H$ satisfy all the statistical QoS requirements. In all cases, the results are compared for different traffic intensities $\lambda$, with Figure (d) yielding the same results for any $\lambda$.}
    \label{fig:simulation_results_partB}
\end{figure}

Figures~\ref{fig:NL_optBH_lambdaBS_lambda} and~\ref{fig:NL_optLoad_lambdaBS_lambda} shows the optimal per-frame bandwidth $B$, and the optimal per-frame computing capacity $H_{f} = \frac{H}{\lambda A}$, and the resulting server load as a function of density of base stations $\lambda_{\mathrm{b}}$, for different traffic intensities $\lambda$. Increasing traffic intensity yields a clear statistical multiplexing gain at the edge servers, reducing the required wireless and computing resources per processed frame. Densification of base stations primarily benefits the wireless link by shortening propagation distances and lowering path loss, thereby reducing the bandwidth needed to satisfy latency constraints. In contrast, denser deployments reduce the average Voronoi cell area and hence the aggregation of arrivals per server, which increases the computing capacity required to maintain tail-delay guarantees. Consequently, the noise-limited regime exhibits a fundamental trade-off: wireless resources benefit from dense deployments, while computational efficiency improves in sparser deployments.

Figure~\ref{fig:NL_optLoad_lambdaBS_lambda} further shows that under low traffic load and high base-station density, servers may operate in an under-utilized regime, increasing the cost per processed frame. In this region, the optimizer compensates by allocating additional bandwidth (despite favorable propagation) to relax the computational constraints imposed by the end-to-end latency requirement. At the same time, notice that all operating points in Figure~\ref{fig:NL_optLoad_lambdaBS_lambda} satisfy the global-optimality condition in Lemma 4.1 based on server load, i.e., $\rho^* \geq \rho_{\min} = 1+W\left(-\omega_{\min}/e\right)$. For $\omega_{\min}=0.8$ (Table~\ref{tab:system_parameters_simulation}), the corresponding minimum load threshold is $\rho_{\min}=0.528$, and the loads remain above this threshold across all configurations. This confirms that the convex reformulation \eqref{eq:optimization_problem_convex} attains the global optimum of the original non-convex dimensioning problem \eqref{eq:optimization_problem_original}.

Figure~\ref{fig:NL_optBH_betaTradeOff} characterizes the optimal dimensioning as a function of the trade-off parameter $\beta_1$ for different traffic intensities $\lambda$ when $\lambda_{\mathrm{b}} = 2$ BS/km$^2$. Varying $\beta_1\in(0,1)$ continuously shifts the resource dimensioning between bandwidth-dominant and compute-dominant provisioning, producing a monotone trade-off between the two resource dimensions. Moreover, since Lemma~\ref{lemma:optimalAnalysis} holds, this parametric sweep generates the complete Pareto frontier of communication versus computation cost per unit area, as stated in Remark~\ref{remark:pareto_frontier}. At the extremes $\beta_1\rightarrow 0$ and $\beta_1\rightarrow 1$, the resource with vanishing weight in the objective can grow large, while the other resource approaches the minimal provisioning required to satisfy the statistical latency and accuracy constraints. The near-symmetry of the curves around $\beta_1=0.5$ corroborates the adequate choice of the scaling parameter $\beta_2$, which is selected to keep the bandwidth and compute terms in \eqref{eq:objective_function_convex} on comparable magnitude.

Finally, Figure~\ref{fig:NL_optrAs_lambdaBS} confirms that optimal values of the auxiliary parameters $r$, $A$, and $s$ are given by their lower bounds, reflecting the monotonicity of the binding constraints. Any user farther from the optimal $r$ would require more wireless resources to avoid large propagation delays; any MEC server serving frames from users within a Voronoi cell area bigger than the optimal $A$ would require more computational resources to avoid overloading the server; and any user offloading frames with any resolution higher than $s$ would require more wireless and computational resources to satisfy the latency requirements. Hence, there no cost-optimal incentive to serve users beyond the selected $r$, to aggregate arrivals over areas larger than $A$, or to offload frames at higher resolution than $s$, since each of these changes strictly tightens the QoS constraints and increases the required provisioning.

\subsection{Interference-limited system}
\begin{figure}[!t]
    \centering
    \subfigure[]{\includegraphics[width=0.44\textwidth]{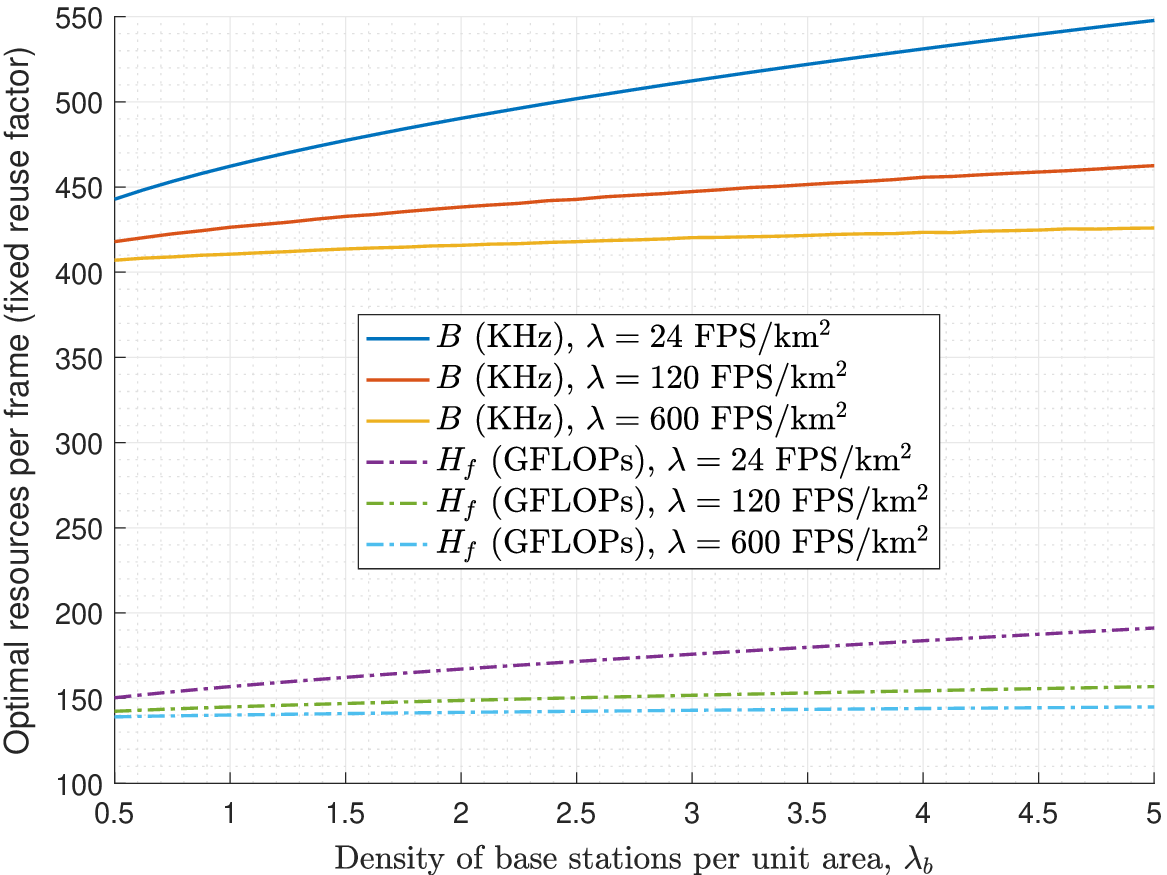} \label{fig:IL_optBH_lambdaBS_lambda_delta4}}$\quad$
    \subfigure[]{\includegraphics[width=0.44\textwidth]{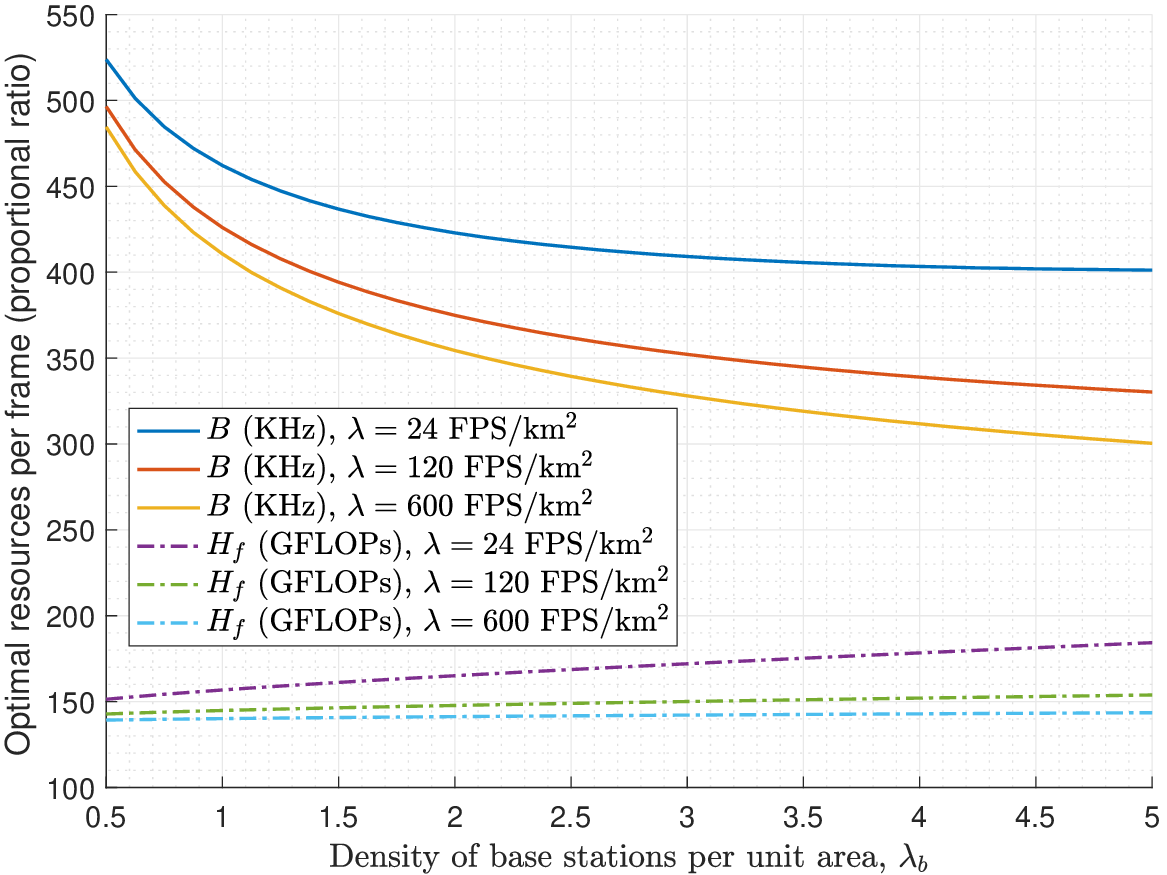} \label{fig:IL_optBH_lambdaBS_lambda_ratio0125}}
    \caption{(a)--(b) Optimal wireless and computing resources per frame for an interference-limited system as a function of the density of base stations per unit area $\lambda_{\mathrm{b}}$. Figures (a) and (b) analyze the effect of having a constant reuse factor $\delta$, or adjusting $\delta$ in proportion to $\lambda_{\mathrm{b}}$, respectively. All the results are also compared for different traffic intensities $\lambda$.}
    \label{fig:simulation_results_partC}
\end{figure}

Figures~\ref{fig:IL_optBH_lambdaBS_lambda_delta4} and~\ref{fig:IL_optBH_lambdaBS_lambda_ratio0125} shows the optimal per-frame bandwidth $B$ and the optimal per-frame computing capacity $H_{f} = \frac{H}{\lambda A}$ in the interference-limited regime as function of the base-station density $\lambda_{\mathrm{b}}$ under two different spectrum-reuse strategies. In either case, we observe again that the communication and computational resources both benefit from statistical multiplexing gain; that is, the higher the traffic in the system, the lower the number of resources required per frame. However, in Figure~\ref{fig:IL_optBH_lambdaBS_lambda_delta4}, the reuse factor $\delta$ is kept constant while the network is densified. In this case, increasing $\lambda_{\rm b}$ leads to a higher density of co-channel interferers, which offsets the benefits of shorter link distances. As a result, the required bandwidth per frame does not decrease with densification. This behavior highlights that densification under fixed reuse is fundamentally inefficient in interference-limited systems, as it simultaneously aggravates interference and reduces server-side aggregation gains.

In contrast, Figure~\ref{fig:IL_optBH_lambdaBS_lambda_ratio0125} considers a proportional reuse strategy in which $\delta$ scales with $\lambda_{\mathrm{b}}$, keeping the ratio $\lambda_{\mathrm{b}}/\delta=0.25$ constant. Under this interference-aware design, densification restores its effectiveness: shorter propagation distances translate directly into reduced transmission delays, allowing both bandwidth and computing resources per frame to decrease with increasing base-station density. The comparison between Figures\ref{fig:IL_optBH_lambdaBS_lambda_delta4} and~\ref{fig:IL_optBH_lambdaBS_lambda_ratio0125} therefore demonstrates that network densification is only cost-efficient in interference-limited regimes when accompanied by appropriate spectrum partitioning. If such adaptation is not considered, denser deployments leads to higher overall provisioning costs and degraded efficiency.

These results emphasize that interference management, either through frequency reuse or equivalent coordination mechanisms, is a first-order design parameter in large-scale edge-intelligent networks. Unlike the noise-limited regime, where densification inherently improves wireless performance, interference-limited systems require coordinated spectrum scaling to unlock the potential benefits of dense deployments.

% Figures~\ref{fig:IL_optBH_lambdaBS_lambda_delta4} and~\ref{fig:IL_optBH_lambdaBS_lambda_ratio0125} shows the optimal per-frame bandwidth $B$ and the optimal per-frame computing capacity $H_{f} = \frac{H}{\lambda A}$ in an interference-limited scenario as a function of density of base stations $\lambda_{\mathrm{b}}$ for different traffic intensities $\lambda$ when $\delta=4$ and when $\lambda_{\mathrm{b}}/\delta=0.25$, respectively. In either case, we observe again that the communication and computational resources both benefit from statistical multiplexing gain; that is, the higher the traffic in the system, the lower the number of resources required per frame. When compared, we observe the importance of selecting an adequate reuse factor $\delta$. For dense networks, the computational resources decay for increasing $\lambda_{\mathrm{b}}$ if and only if $\delta$ is adjusted proportionally to $\lambda_{\mathrm{b}}$, otherwise the computational resources must be increased to cope with the increasing interference. Contrarily, sparse networks benefit from setting a fixed $\delta$ because it prevents the network from being over-dimensioned.

\section{Conclusions}
\label{sec:conclusions}
This work established a comprehensive stochastic framework for the dimensioning of wireless and computational resources in large-scale multi-cellular edge-intelligent systems. By jointly modeling spatial network randomness, uplink transmission dynamics, queueing behavior at edge servers, and AI inference workloads, we characterized the end-to-end performance of edge video analytics under realistic deployment conditions. Unlike prior work that focuses primarily on run-time adaptation or average performance metrics, our approach addresses the planning-stage problem of resource provisioning under strict statistical guarantees on latency, coverage, and inference accuracy.

A key analytical contribution is the derivation of closed-form expressions for the ergodic uplink capacity in both noise-limited and interference-limited regimes, explicitly accounting for Poisson–Voronoi cell geometry, fractional power control with peak-power constraints, and frequency reuse. These results expose how base-station density, reuse factor, power-control parameters, and receiver antenna count jointly shape throughput, fairness, and interference. In particular, we showed that densification of cellular radio access network alone does not universally improve performance: while shorter link distances benefit noise-limited systems, interference-limited networks can experience increased throughput disparity and higher provisioning costs unless interference is explicitly managed.

Building on end-to-end offloading model, we formulated a global optimization problem that jointly dimensions bandwidth and edge-computing hardware to minimize deployment cost while satisfying statistical guarantees on latency and inference accuracy. In particular, the inherent relationship between the wireless and computing resources is captured as a function of the supported arrival rate of tasks at the edge server, the maximum permitted end-to-end delay in the system, the density of base stations in the network, and two intrinsic methods to manage the inter-cell interference in large-scale networks: the cellular reuse factor and the per-user power control. Despite the inherent non-convexity of the joint optimization, we provided a theoretical analysis to demonstrate that the problem can be decomposed into a sequence of convex sub-problems, guaranteeing zero optimality gap under specific load conditions. This result enables practical and reliable optimization of large-scale edge deployments without resorting to heuristic approximations.

Our numerical results reveal fundamental trade-offs between wireless and computational efficiencies in network design. In noise-limited regimes, wireless resources benefit from network densification due to reduced path loss, whereas computational efficiency improves in sparser deployments that allow greater statistical multiplexing at edge servers. In interference-limited regimes, however, densification is only cost-efficient when accompanied by proportional scaling of the frequency reuse factor; otherwise, fixed reuse leads to excessive interference, reduced fairness among cell-edge users, and increased computational over-provisioning. These findings underscore that interference management is a first-order design parameter for edge intelligence and must be considered jointly with compute provisioning.

Overall, this work demonstrates that cost-effective edge intelligence cannot be achieved by dimensioning radio access and edge computing in isolation. Instead, it requires a holistic, stochastic design methodology that captures spatial variability, traffic aggregation, and workload characteristics. Moreover, our proposed framework provides network operators and system designers with quantitative guidelines for balancing spectrum allocation, hardware investment, and quality-of-service guarantees in future 5G and beyond-6G edge-intelligent networks.

\bibliography{references}
\bibliographystyle{IEEEtran}

\appendix
\subsection{Ergodic capacity of noise-limited systems}
\label{app:erg_NL}
Recall the uplink SINR expression in~\eqref{eq:SINR}. In a noise-limited regime, where inter-cell interference is negligible, the interference term in the denominator of~\eqref{eq:SINR} vanishes and
the SINR reduces to the signal-to-noise ratio (SNR):
\begin{equation}
    \mathrm{SNR} = \frac{\|\bm{g}\|^2_2 \,\ell(r,\alpha,\epsilon)\,r^{-\alpha}}{\sigma^2}.
    \label{eq:SINR_noise_limited}
\end{equation}

For a user with allocated bandwidth $B$ transmitting at distance $r$ from the base station, all quantities in \eqref{eq:SINR_noise_limited} are deterministic except for the small-scale fading gain $\lVert \bm{g} \rVert_2^2$. Hence, the only source of randomness in the SNR is the channel gain. In this setting, the ergodic capacity for such a user can be calculated as
\begin{align*}
    \Bar{C}_{\mathrm{NL}}(B,r) & = \mathbb{E}[B \log_2(1+\text{SNR})] \nonumber \\
    & \overset{(a)}{=} \frac{B}{\log(2)} \int_{0}^{\infty} \mathcal{P}\left(\log(1+\text{SNR})\geq t\right)dt \nonumber \\
    %& = \frac{B}{\log(2)} \int_{0}^{\infty} \mathcal{P}\left(\text{SNR}\geq e^{t}-1\right)dt \nonumber \\
    & \overset{(b)}{=} \frac{B}{\log(2)} \int_{0}^{\infty} \mathcal{P}\left( \|\bm{g}\|^2_2 \geq \frac{(e^t-1)\, \sigma^2}{\ell(r,\alpha,\epsilon)\,r^{-\alpha}} \right)dt \nonumber
    %& \overset{(c)}{=} \frac{B}{\log(2)} \int_{0}^{\infty} \exp\left(- \frac{(e^t-1)\, r^{\alpha}}{\gamma\,\ell(r,\alpha,\epsilon)}\sigma^2\right) dt \nonumber \\
    %& \overset{(d)}{=} \frac{B}{\log(2)} \, \exp\left(\frac{B N_0\,r^{\alpha}}{\gamma\,\ell(r,\alpha,\epsilon)}\right) \, E_1\left(\frac{B N_0\,r^{\alpha}}{\gamma\,\ell(r,\alpha,\epsilon)}
    %\right),
\end{align*}
where $(a)$ follows from the fact that the expectation of any positive random variable $X$ can be calculated in the sense of Lebesgue-Stieltjes as $\mathbb{E}[X]=\int_0^\infty \mathcal{P}\left(X\geq t\right)dt$, and $(b)$ follows from the definition of the SNR in~\eqref{eq:SINR_noise_limited}. If we further define the variable $\theta = \frac{\sigma^2 r^\alpha}{\gamma \ell(r,\alpha,\epsilon)}$, make the substitution $y=(e^t-1)\theta$, and consider that $\|\bm{g}\|^2_2$ follows the Gamma distribution
\begin{equation}
    \mathcal{P}(\|\bm{g}\|^2 \leq x) = 1 - \exp\left(-\frac{x}{\gamma}\right) \sum_{i=0}^{M-1} \frac{1}{i!} \left(\frac{x}{\gamma}\right)^i, \quad x > 0,
    \label{eq:Gamma_channel_gain}
\end{equation}
we can express the ergodic capacity as
\begin{align}
    \Bar{C}_{\mathrm{NL}}(B,r) & = \frac{B}{\log(2)} \sum_{i=0}^{M-1} \frac{1}{i!} \int_{0}^{\infty} \frac{y^i}{y+\theta} e^{-y} dy \nonumber \\
    & \overset{(d)}{=} \frac{B}{\log(2)} \sum_{i=0}^{M-1} \frac{1}{i!} \left[\int_{0}^{\infty} y^{i-1}e^{-y} dy - \theta\int_{0}^{\infty} \frac{y^{i-1}}{y+\theta}e^{-y}dy\right] \nonumber \\
    & \overset{(e)}{=} \frac{B}{\log(2)} \sum_{i=0}^{M-1} e^\theta E_{i+1}(\theta)\nonumber
\end{align}
where $(d)$ follows from partial fraction decomposition, and $(e)$ follows from iteratively applying the same principle as in $(d)$  and using the fact that the generalized exponential integral function, defined as $E_i(x) = \int_1^\infty \frac{e^{-xt}}{t^i}\, dt$, satisfies the recursion
\begin{equation*}
    E_{i+1}(x)=\frac{1}{i}\left[e^{-x}-xE_i(x)\right], \qquad \forall i\geq1.
\end{equation*} %Recall also that the Gamma function, defined as $\Gamma(i)=\int_0^\infty y^{i-1}e^{-y}$, is simply $(i-1)!$ when $i$ is a natural number.

Lastly, after substituting back the expression for $\theta$ and expressing the noise power $\sigma^2$ as the product of the power spectral density of the noise $N_0$ and the allocated bandwidth $B$, the final closed-form solution of the ergodic capacity results in
\begin{equation*}
    \Bar{C}_{\mathrm{NL}}(B,r) = \frac{B}{\log(2)} \exp\left(\frac{BN_0 \,r^\alpha}{\gamma \ell(r,\alpha,\epsilon)} \right) \sum_{i=0}^{M-1} E_{i+1}\left(\frac{BN_0 \, r^\alpha}{\gamma \ell(r,\alpha,\epsilon)} \right).
\end{equation*}

\subsection{Ergodic capacity of interference-limited systems}
\label{app:erg_IL}
Starting from the uplink SINR expression in~\eqref{eq:SINR}, the interference-limited regime corresponds to neglecting thermal noise in the denominator. The SINR therefore reduces
to the signal-to-interference ratio (SIR):
\begin{equation}
    \mathrm{SIR} = \frac{\|\bm{g}\|^2_2 \,\ell(r,\alpha,\epsilon)\,r^{-\alpha}}{\sum_{z\in\mathcal{Z}} \|\Pi_{\bm{g}}\,\bm{g}_z\|^2_2\, \ell(r_z,\alpha,\epsilon)\, d_z^{-\alpha}}.
    \label{eq:SINR_interference_limited}
\end{equation}

For a user with allocated bandwidth $B$ transmitting at distance $r$ from the base station, the sources of randomness in the SIR in~\eqref{eq:SINR_interference_limited} are the channel coefficients, the distances between the interfering users and their serving base stations, and the distances between the interfering users and our base station of interest. The ergodic capacity can then be calculated as
\begin{align*}
    \Bar{C}_{\mathrm{IL}}(B,r) & = \mathbb{E}[B \log_2(1+\text{SIR})] \\
    & \overset{(a)}{=} \frac{B}{\log(2)}\, \mathbb{E}\left[\int_{0}^{\infty} \frac{1}{s}\left(1-e^{-s\|\bm{g}\|^2_2\,\ell(r,\alpha,\epsilon)\,r^{-\alpha}}\right)\exp\left( -s \sum_{z\in\mathcal{Z}} \|\Pi_{\bm{g}}\,\bm{g}_z\|^2_2\, \ell(r_z,\alpha,\epsilon)\, d_z^{-\alpha}\right) ds\right] \\
    & \overset{(b)}{=} \frac{B}{\log(2)}\int_{0}^{\infty}  \frac{1}{s}\left(1-\mathbb{E}\left[e^{-s\|\bm{g}\|^2_2\,\ell(r,\alpha,\epsilon)\,r^{-\alpha}}\right]\right)\, \mathbb{E}\left[\exp\left( -s \sum_{z\in\mathcal{Z}} \|\Pi_{\bm{g}}\,\bm{g}_z\|^2_2\, \ell(r_z,\alpha,\epsilon)\, d_z^{-\alpha}\right)\right] ds \\
    & \overset{(c)}{=} \frac{B}{\log(2)}\, \int_{0}^{\infty} \frac{1}{s}\Big(1-(1+s\,\gamma \ell(r,\alpha,\epsilon)\,r^{-\alpha})^{-M}\Big) \mathcal{L}(s) ds,
\end{align*}
where $(a)$ follows from the proof in~\cite[Lemma 1]{hamdi2010useful} and the definition of the SIR, $(b)$ follows from the independence between the signal of interest and the interfering signals, and $(c)$ follows from the Laplace transform of the Gamma distribution of $\|\bm{g}\|^2_2$ in~\eqref{eq:Gamma_channel_gain} and the definition of the Laplace transform of the interference, $\mathcal{L}(s)$.

Now, to derive $\mathcal{L}(s)$, we need to calculate Laplace transform of the aggregate interference experienced in the serving base station of the user of interest. For that, recall that the set $\mathcal{Z}$ is the point process of co-channel interferers. Given the frequency reuse factor $\delta$, the effective density of co-channel base stations is $\lambda_b / \delta$. The locations of the interfering users are displaced from their serving base stations by distances $r_z$, considered here to be Rayleigh distributed. Moreover, using the approximation that the $r_z$ are i.i.d. and independent of $d_z$, and modeling the interfering base station locations as a PPP with density $\lambda_b / \delta$ outside the guard radius $r$ (since the serving base station is the nearest, there can be no interferers closer than $r$), we have:
\begin{align*}
    \mathcal{L}(s) & = \mathbb{E}\left[\exp\left( -s \sum_{z\in\mathcal{Z}} \|\Pi_{\bm{g}}\,\bm{g}_z\|^2_2\, \ell(r_z,\alpha,\epsilon)\, d_z^{-\alpha}\right)\right] \\
    & = \mathbb{E}\left[\,\prod_{z\in\mathcal{Z}} \exp\left( -s \|\Pi_{\bm{g}}\,\bm{g}_z\|^2_2\, \ell(r_z,\alpha,\epsilon)\, d_z^{-\alpha}\right)\,\right] \\
    & \overset{(d)}{=} \mathbb{E}\left[\,\prod_{z\in\mathcal{Z}}\, \int_{0}^{\infty} 2\pi\lambda_{\mathrm{b}}u\, e^{-\pi\lambda_{\mathrm{b}} u^2} \exp\left( -s \|\Pi_{\bm{g}}\,\bm{g}_z\|^2_2\, \ell(u,\alpha,\epsilon)\, d_z^{-\alpha}\right)\,du\,\right] \\
    & \overset{(e)}{=} \mathbb{E}\left[\,\prod_{z\in\mathcal{Z}}\, \int_{0}^{\infty}\frac{2\pi\lambda_{\mathrm{b}}u\, e^{-\pi\lambda_{\mathrm{b}} u^2}}{1+s\,\gamma \ell\left(u,\alpha,\epsilon\right) d_z^{-\alpha}}\,du\,\right] \\
    & \overset{(f)}{=} \exp\left(-2\pi\frac{\lambda_{\mathrm{b}}}{\delta}\int_{r}^{\infty} \left(1-\int_{0}^{\infty}\frac{2\pi\lambda_{\mathrm{b}}u\, e^{-\pi\lambda_{\mathrm{b}} u^2}}{1+s\,\gamma \ell\left(u,\alpha,\epsilon\right) x^{-\alpha} }\,du\right)\, x\, dx  \right),
\end{align*}
where $(d)$ follows from the i.i.d. Rayleigh distributions of $r_z$, $(e)$ follows from the Laplace transform of the Exponential distribution of $\|\Pi_{\bm{g}}\,\bm{g}_z\|^2_2$, and $(f)$ follows from the probability generating function of the PPP that models the location of the interfering users~\cite{chiu2013stochastic} when considering polar coordinates. 

To ease the notation, we define the function
\begin{equation*}
    \beta(x, s) = 1 - \int_{0}^{\infty}\frac{2\pi\lambda_{\mathrm{b}}u\, e^{-\pi\lambda_{\mathrm{b}} u^2}}{1+s\,\gamma \ell\left(u,\alpha,\epsilon\right) x^{-\alpha} }\,du
\end{equation*}
to represent the probability that an interferer at distance $x$ from the serving base station contributes to the aggregate interference, with the integral taken over the Rayleigh-distributed displacement $r_z = u$ of the interfering users from their own base station. This compact form leads directly to the expression for $\mathcal{L}(s)$ as stated in Lemma~\ref{lemma:interference_limited}.

\end{document}